\documentclass[twocolumn,aps,pra,longbibliography,superscriptaddress,nofootinbib,10pt]{revtex4-1}
\usepackage[latin1]{inputenc}
\usepackage{graphicx,dcolumn,bm}
\usepackage{subfigure}
\usepackage{multirow}
\usepackage{array}
\usepackage{arydshln}
\usepackage{color}
\usepackage[colorlinks,linkcolor=blue,citecolor=blue,hyperindex]{hyperref}
\usepackage{amsfonts}
\usepackage{amssymb}
\usepackage{amsmath}
\usepackage{latexsym}
\usepackage{simplewick}
\usepackage{times,txfonts}
\usepackage{epstopdf}
\usepackage{multirow}
\usepackage{float}
\usepackage[table]{xcolor}
\usepackage{natbib}
\usepackage{mathrsfs}
\usepackage[sans]{dsfont}
\usepackage[mathscr]{eucal}

\usepackage[normalem]{ulem}
\usepackage{epsfig}
\definecolor{Blue}{rgb}{0.0,0.0,1}
\definecolor{Red}{rgb}{1,0.0,0.0}
\definecolor{Green}{rgb}{0,0.5,0.0}
\usepackage{tikz}
\usetikzlibrary{decorations.pathmorphing}
\usetikzlibrary{arrows}
\usetikzlibrary{intersections,shapes.arrows}
\usetikzlibrary{calc}
\usetikzlibrary{quotes,angles}
\usepackage{nicefrac}
\usepackage{pgfplots}
\usepgfplotslibrary{fillbetween}
\pgfplotsset{compat=1.13,colormap={violetnew}{rgb=(0.293416, 0.0574044, 0.529412) rgb=(0.394818,0.233715,0.671945) rgb =(0.49622,0.410025,0.814477) rgb=(0.588672,0.567494,0.910066) rgb=(0.663226,0.687282,0.911765) rgb=(0.73778,0.807069,0.913465) rgb=(0.807267,0.861883,0.894034) rgb=(0.874222,0.884211,0.864039) rgb=(0.941176, 0.906538, 0.834043)}}
\usepgfplotslibrary{groupplots} 
\usepgfplotslibrary[groupplots] 
\usetikzlibrary{pgfplots.groupplots} 
\usetikzlibrary[pgfplots.groupplots] 
\usepgfplotslibrary{statistics}
\usepackage{pgfplotstable}
\tikzset{jumpdot/.style={mark=*,solid},excl/.append style={jumpdot,fill=white},incl/.append style={jumpdot,fill=black}}

\begin{document}

\title{Experimental Investigation of Geometric Quantum Speed Limits in an Open Quantum System}
\author{Diego Paiva Pires}
\affiliation{Departamento de F\'{i}sica, Universidade Federal do Maranh\~{a}o, Campus Universit\'{a}rio do Bacanga, 65080-805, S\~{a}o Lu\'{i}s, Maranh\~{a}o, Brazil}
\author{Eduardo R. deAzevedo}
\affiliation{Instituto de F\'{i}sica de S\~{a}o  Carlos, Universidade de S\~{a}o Paulo, CP 369, 13560-970, S\~{a}o Carlos, S\~{a}o Paulo, Brazil}
\author{Diogo O. Soares-Pinto}
\affiliation{Instituto de F\'{i}sica de S\~{a}o  Carlos, Universidade de S\~{a}o Paulo, CP 369, 13560-970, S\~{a}o Carlos, S\~{a}o Paulo, Brazil}
\author{Frederico Brito}
\affiliation{Quantum Research Center, Technology Innovation Institute, P.O. Box 9639, Abu Dhabi, UAE}
\affiliation{Instituto de F\'{i}sica de S\~{a}o  Carlos, Universidade de S\~{a}o Paulo, CP 369, 13560-970, S\~{a}o Carlos, S\~{a}o Paulo, Brazil}
\author{Jefferson G. Filgueiras}
\affiliation{Instituto de Qu\'{i}mica, Universidade Federal Fluminense, Outeiro de S\~{a}o Jo\~{a}o Batista, s/no, Niter\'{o}i, 24020-141 RJ, Brazil}
\affiliation{Instituto de F\'{i}sica de S\~{a}o  Carlos, Universidade de S\~{a}o Paulo, CP 369, 13560-970, S\~{a}o Carlos, S\~{a}o Paulo, Brazil}
\affiliation{Instituto de F\'{i}sica, Universidade Federal do Rio de Janeiro, CP 68528, 21941-972, Rio de Janeiro, RJ, Brazil}

\begin{abstract}
We studied geometric quantum speed limits (QSL) of a qubit subject to decoherence in an ensemble of chloroform molecules in a Nuclear Magnetic Resonance experiment. The QSL is a fundamental lower bound on the evolution time for quantum systems undergoing general physical processes. To do so, we controlled the system-reservoir interaction and the spin relaxation rates by adding a paramagnetic salt, which allowed us to observe both Markovian and non-Markovian open system dynamics for the qubit. We used two distinguishability measures of quantum states to assess the speed of the qubit evolution: the quantum Fisher information (QFI) and Wigner-Yanase skew information (WY). For non-Markovian dynamics and low salt concentrations, we observed crossovers between QSLs related to the QFI and WY metrics. The WY metric sets the tighter QSL for high concentrations and Markovian dynamics. We also show that QSLs are sensitive even to small fluctuations in spin magnetization.
\end{abstract}

\maketitle



\textit{\textbf{Introduction}} --- One of the core concepts of quantum mechanics is the uncertainty principle. While this relationship is well known for non-commuting observables, e.g., position and momentum, the time-energy uncertainty relation has been controversial over decades, resulting in several attempts to address this issue~\cite{PhysRev.122.1649,RevModPhys.67.759}. In their seminal work, Mandelstam and Tamm (MT)~\cite{Mandelstam} reinterpreted this question by introducing the concept of quantum speed limit (QSL), which is a threshold imposed by quantum mechanics to the minimum evolution time between two orthogonal states. In this setting, Margolus and Levitin (ML)~\cite{1992_PhysicaD_120_188} derived a bound for the orthogonalization time of pure quantum states that scales with the inverse of the mean energy of the system.

Over a decade ago, Taddei \emph{et al.}~\cite{Taddei2013} have presented a general QSL bound based on the quantum Fisher information, valid for both unitary and nonunitary evolutions. At the same time, del Campo \textit{et al.}~\cite{delcampo2013} obtained a QSL relying on the relative purity, finding evidence that the spectral property of the noise would have an influence on the speed of evolution. In turn, Deffner and Lutz~\cite{PhysRevLett.111.010402} introduced a QSL bound in terms of the operator norm of the generator of the nonunitary dynamics. Such an approach allowed them to investigate the influence of non-Markovianity on the evolution rate, finding that, for the Jaynes-Cummings model, its effect could lead to a faster evolution. QSLs have been addressed for either closed and open quantum systems~\cite{PhysRevLett.120.070402,Quantum_3_168_2019,PhysRevA.103.022210,JPhysAMathTheor_54_395304_2021,PhysRevA.104.052620,arXiv:2110.13193,arXiv:2004.03078,PhysRevA.106.012403,PhysRevLett.129.140403}, and find applications ranging from quantum many-body systems~\cite{PhysRevLett.124.110601,PhysRevResearch.2.023299,PhysRevResearch.2.032020,PhysRevA.102.042606,PhysRevX.11.011035,PhysRevLett.126.180603,arXiv:2006.14523}, to quantum thermodynamics~\cite{PhysRevE.103.032105,PhysRevA.104.052223,arXiv:2204.10368,arXiv:2203.12421}. In the context of nonequilibrium quantum dynamics, Ref.~\cite{IntJModPhysB_36_2230007_2022} provides an extensive review on quantum speed limit bounds of several dynamical evolutions, particularly focusing on quantum many-body systems.

Information geometry is a powerful tool to study QSLs. In this setting, one finds a general framework providing an infinite family of QSLs based on contractive Riemannian metrics on the space of quantum states, which applies to any physical process~\cite{Diego2017}. In this scenario, the QSL relates to a certain information-theoretic distinguishability measure built for a set of quantum states, pure or mixed, separable or entangled, valid for closed and open systems. In addition, QSL bounds were investigated theoretically using matrix norms, e.g., Schatten $p$-norms, applied to the generator of the nonunitary dynamics~\cite{Deffner_2017,NewJPhys_19_103018_2017,arXiv:2312.00533,arXiv:2401.01746}. Recent results include the study of bounds on the speed of observables related to open quantum systems~\cite{PhysRevX.12.011038}, and also the proposal of a general framework for deriving tighter speed limits for macroscopic systems~\cite{PRXQuantum.3.020319}, which in turn finds applications in transport phenomena and nonequilibrium dynamics in spin systems.

In the last years, several works include an analysis of QSLs for driven quantum systems under Markovian evolution~\cite{PhysRevLett.111.010402,NJPhys_24_055003_2022}. These results triggered studies of QSLs for particular choices of quantum channels focusing on how the degree of non-Markovianity could affect the tightness of QSL bounds~\cite{Xu2014,Sun2015,Meng2015,Nicolas2016,Zhang2016}. Overall, non-Markovian dynamics present an intricate physical structure, allowing the revivals of genuine quantum resources during the evolution, for example, quantum coherence, thus contrasting with the typical monotonic loss observed in Markovian scenarios~\cite{Rivas_2014}. In Ref.~\cite{SciRep_6_38149_2016}, the authors discussed timescales related to generate an amount of quantumness under arbitrary physical process, also investigating the dependence of the QSL bound on initial and final states, and the role played by non-Markovian effects. Noteworthy, QSL depends on the initial state and the dynamical map go\-verning the system evolution, showing the absence of a general connection between non-Markovianity and the speed of evolution of a quantum system~\cite{PhysRevA.89.012307,Cianciaruso2017,Teittinen2019,Entropy_23_331_2021}.

Recently, an experimental measurement of the QSL time in a trapped single-atom system showed a crossover between the Mandelstam-Tamm and Margolus-Levitin bounds, with the latter dominating the dynamics for longer times~\cite{sciadvabj91192021}. We also mention the experimental discussion of the MT bound for time-dependent Hamiltonians with nuclear spin systems~\cite{PhysRevA.97.052125}. Despite all theoretical advances in understanding QSLs and experimental achievements for closed quantum systems, the field still lacks experimental studies exploring and certifying the machinery developed for open quantum systems. That happens due to the challenge of controlling the system-environment interaction and configuration, which determines the Markovian or non-Markovian character of the evolution.

In this work, we assess geometric QSLs in an open quantum system, by controlling just two parameters of the bath: the relaxation times of the hydrogen and carbon nuclear spins of an ensemble of chloroform molecules in a liquid-state Nuclear Magnetic Resonance (NMR) experiment. We control the relaxation rates of the carbon and hydrogen nuclear spins by adding a paramagnetic salt to the solution, allowing us to observe the transition from non-Markovian to Markovian regimes in the dynamics~\cite{KondoNJP,Kondo2020}. Then, we investigate how the speed of evolution is affected under different experimental conditions. For non-Markovian dynamics and low salt concentrations, we observe crossovers between quantum speed limits defined by the quantum Fisher information and the Wigner-Yanase skew information metrics. The occurrence of these crossovers is related to the character of the systems time evolution. In high concentrations, the Wigner-Yanase metric sets the tighter quantum speed limit bound when the system undergoes non-Markovian or Markovian dynamics.




\textit{\textbf{Model}} --- In our experiments, we consider a two-qubit system encoded on the nuclear spins of the $^{13}$C and $^{1}$H of an ensemble of chloroform (CHCl$_3$) molecules in liquid-state at room temperature. The interaction between the spins is given by the Hamiltonian $H = \frac{\pi J}{2} {\sigma_z}\otimes {\sigma_z}$, where $J$ is the strength of the interaction, and $\sigma_{i}$ is the $i$-th Pauli matrix ($i = x, y, z$). For simplicity, we set $\hbar = 1$. The two-qubit system initiates in a thermal equilibrium state ${\rho^{CH}_T} \approx [(1 - {\epsilon_C})/4] \, \mathbb{I}\otimes\mathbb{I}  + ({\epsilon_C}/2) \, |0\rangle\langle{0}|\otimes\mathbb{I} + ({\epsilon_H}/4) \, \mathbb{I}\otimes{\sigma_z}$, which is valid at the high-temperature limit ($\epsilon_l\ll1$), with ${\epsilon_l} = \hbar{\omega_l}/{k_B}T$ and ${\omega_{C,H}}$ is the $^{13}$C ($^{1}$H) Larmor frequency, ${k_B}$ is the Boltzmann constant, and $T$ is the temperature. Since we are interested in the $^{13}$C magnetization, one can discard the first and third terms in state ${\rho^{CH}_T}$, such that the density matrix for the $^{13}$C spin is given, up to normalization, by ${\rho^C} = {\text{Tr}_H}(\,{\rho_T^{CH}}) \approx |{0}\rangle\langle{0}|$~\cite{KondoNJP}. We obtain the initial state for our experiments after a $\pi/4$ rotation on the $y$-axis, resulting in ${\rho_0^C} = (1/2)(\mathbb{I} + ({\sigma_x} + {\sigma_z})/\sqrt{2})$.

To model the open system dynamics of the carbon nuclear spin, we followed the description in Refs.~\cite{KondoNJP, KondoPRA}. The decoherence for each nuclear spin is described by the longitudinal and transverse relaxation times, $T_{1,(C,H)}$ and $T_{2,(C,H)}$, respectively. The longitudinal relaxation time is associated to the spin-lattice relaxation, i.e., the return of the system to thermal equilibrium. The transverse relaxation time describes the spin-spin relaxation, which affects only the coherences of the density matrix.

In our system, the $^{1}$H spin becomes a source of the decoherence for the carbon spin when the condition ${T_{1,H}} \approx 1/J$ is satisfied. Under this condition, the flips on the proton spin due to ${T_{1,H}}$ occur at the same time scale as the evolution caused by the scalar coupling, introducing an effective time dependence on this coupling. This results in a faster decay of the system magnetization, known as scalar relaxation~\cite{abragam}. This process is inhomogeneous in time, with a correlation time given by the minimum between ${T_{1,H}}$ and $1/J$~\cite{KondoNJP, KondoPRA, Ho_2019}. By modulating ${T_{1,H}}$ through paramagnetic relaxation, we can control the degree of non-Markovianity of the $^{13}$C open-system dynamics. In the limit of fast correlation time, ${T_{1,H}} \ll 1/J$, i.e., for high concentrations of the paramagnetic salt, we observe Markovian dynamics for the system dephasing.

We consider a phase-damping channel to model the spin-spin relaxation of the $^{13}$C spin and a single bit-phase flip channel for both spin-lattice and spin-spin relaxations of the $^{1}$H spin, since ${T_{1,H}} \approx {T_{2,H}}$ for all experimental configurations discussed throughout our results. The Kraus operators for the phase damping are ${K_1} = \sqrt{q_{t}} \, \mathbb{I}\otimes\mathbb{I}$ and ${K_2} = \sqrt{1 - {q_{t}}} \, {\sigma_z} \otimes \mathbb{I}$, with ${q_t} = (1 + {e^{-t/2{T_{2,C}}}})/2$. For the bit-phase flip, one gets ${\mathcal{E}_1} = \sqrt{{p_t}} \, \mathbb{I}\otimes\mathbb{I}$ and ${\mathcal{E}_2} = \sqrt{1 - {p_{t}}} \, \mathbb{I} \otimes {\sigma_y}$, where ${p_t} = (1 + {e^{-t/2{T_{1,H}} } } )/2$. Here ${T_{1,H}}$ and ${T_{2,C}}$ define the characteristic time of the bit-phase flip channel and phase damping channel, respectively. 
 
In this scenario, the state of the two-qubit system during the evolution is obtained by the application of the Kraus operators to the initial state, when partitioning the evolution time $0 \leq t \leq \tau$ into $N$ equal steps of size $\Delta t = \tau/N$, with $J\Delta t \ll 1$. Hence, by iterating such a process, with $\Delta t \rightarrow 0$ ($N \rightarrow \infty$), and tracing out the hydrogen degrees of freedom, one obtains the single-qubit marginal state as follows
\begin{equation}
\label{eq:000000000001}
{\rho_t^C} = \frac{1}{2}\left(\mathbb{I} + {\langle{\sigma_x}\rangle_t}\,{\sigma_x} + {\langle{\sigma_z}\rangle_0} \, {\sigma_z} \right) ~,
\end{equation}
where ${\langle{\sigma_x}\rangle_t} = {\xi(t)}\, {\langle{\sigma_x}\rangle_0}$ stands for the transversal magnetization of the $^{13}$C (hereafter the system), with 
\begin{align}
\label{eq:000000000002}
{\xi(t}) &= {e^{-t/2{T_{2,C}}}}{e^{-t/4{T_{1,H}}}}\left[\frac{t}{4{T_{1,H}}} \text{sinc}\left(\frac{t}{4{T_{1,H}}}\sqrt{16 \, {\pi^2}{J^2}{T_{1,H}^2} - 1} \, \right) \right.\nonumber\\
&\left.+ \text{cos}\left(\frac{t}{4{T_{1,H}}}\sqrt{16 \, {\pi^2}{J^2}{T_{1,H}^2} - 1}\, \right)  \right] ~,
\end{align}
with $\text{sinc}(x) := \sin{x}/x$, and $\xi(0) = 1$. For more details, see the Supplementary Materials~\cite{SupMat}. From now on, we will omit the superscript in Eq.~\eqref{eq:000000000001} and define $\rho_t$ as the single-qubit state of the $^{13}$C nuclear spin.

The function $\xi(t)$ in Eq.~\eqref{eq:000000000002} encodes the nonunitary signatures in the dynamics of the two-level system. Furthermore, it shows how reducing ${T_{1,H}}$ increases the role of the $^{1}$H nuclear spin as the main source of decoherence, with the dynamics dominated by the $\text{sinc}(\bullet)$ term in Eq.~\eqref{eq:000000000002} and the oscillations due to the scalar interaction disappearing for ${T_{1,H}} \ll 1/J$. Thus, to evaluate the geometric QSL of a given Riemannian metrics, we prepare the state $\rho_0$ by the application of a single $(\pi/4)$ pulse on the y-axis and let the system evolve according to Eq.~\eqref{eq:000000000002}. We control the relaxation rates $1/{T_{1,H}}$ and $1/{T_{2,C}}$ by adding the paramagnetic salt iron(III) acetylacetonate (Fe(acac)$_3$) to the solution, with the relaxation rates growing linearly with the concentration of Fe(acac)$_3$~\cite{abragam}.



\textit{\textbf{Methods}} --- We carried out the measurements at 25$^o$ C in a Bruker Avance III 600 MHz, with $^{1}$H and $^{13}$C Larmor frequencies of 600 and 150 MHz, respectively, with a 5 mm double resonance probe-head. We realized the experiments with a solution of chloroform (CHCl$_{3}$), with natural abundance of $^{13}$C in a 5 mm NMR tube, doped with Iron(III) acetylacetonate (Fe(acac)$_3$, Sigma Aldrich). To guarantee good frequency stability, we used the deuterium signal of acetone-d6 (Cambridge Isotopes Laboratories - Inc.) to lock the NMR signal. We avoided the undesired line broadening of the deuterium reference signal, due to the effects of the paramagnetic salt, putting the acetone-d6 in a 3 mm NMR tube, all inside the 5 mm tube.

We prepared the solutions of CHCl$_3$ and Iron(III) acetylacetonate diluting the paramagnetic salt in 2 ml of CHCl$_3$. To get the concentrations of 20, 50, 120, 300 and 450 mM, we used 14.5(1), 35.5(1), 85.1(1), 211.6(1) and  317.8(1) miligrams of Fe(acac)$_3$, respectively. The error on the concentrations is of 1 mM. Each sample contained 150 $\mu$l of doped CHCl$_3$ and 150 $\mu$l of acetone-d6.

We measured the spin-lattice relaxation time ${T_{1,(H, C)}}$ for $^{1}$H and $^{13}$C using a standard inversion-recovery pulse sequence. We estimate $T_{1,(H,C)}$ adjusting the time delay $t_1$ in the sequence such that the magnetization vanishes. Under this condition, we can calculate $T_{1,(H,C)}$ from the equation $T_1 = t_1/\ln(2)$. The $^{13}$C spin-spin relaxation time ${T_{2,C}}$ is measured from the Free Induction Decay (FID) signal obtained when both spins are decoupled through a Waltz-64 heteronuclear decoupling sequence \cite{Waltz64}, with a decoupling $\pi/2$-pulse of 54 $\mu$s. We assumed ${T_{2,C}} \approx T_{2,C}^{*}$ due to a good shimming. Here, $T_{2,C}^{*}$ is the characteristic time for the FID decay and it differs from $T_{2,C}$ due to the effects of inhomogeneities in the static field $B_0$~\cite{abragam}. To avoid frequency offset effects on the estimation of $T_{2,C}$, the FID signals were fitted using a function ${M_x}(t) = {M_0} \, {e^{- t/{T_{2,C}}}}\cos(\omega t)$, where $M_0$ is the initial magnetization amplitude and $\omega$ is the frequency offset. We used the spectrum of the thermal equilibrium state to adjust the phase and normalize the intensity of all NMR data throughout this work.



\textit{\textbf{Quantum Speed limits for Open System Dynamics}} --- The quantum speed limit (QSL) is related to the distinguishability of quantum states from a geometric perspective~\cite{Diego2017}. We remind that the so-called Morozova-\v{C}encov-Petz (MCP) theorem states that the convex space of quantum states is endowed with a family of contractive Riemannian metrics~\cite{MC-paper,Petz-paper}. In this sense, the nonuniqueness of distinguishability measures of quantum states implies a class of geometric QSLs that can be exploited in the search for tighter bounds. Here, we discuss the QSL time for the single-qubit state ${\rho_t}$ in Eq.~\eqref{eq:000000000001}. The evolution of this state draws a path $\gamma$ in the space of quantum states connecting initial $\rho_0$ and final $\rho_{\tau}$ states. The MCP theorem states that the length ${\ell^f_{\gamma}}({\rho_{0}},{\rho_{\tau}})$ of such path depends on some chosen contractive Riemannian metric related to a given Morozova-\v{C}encov (MC) function $f$~\cite{MC-paper,Petz-paper}. Hereafter, we will restrict our analysis to two paradigmatic metrics: the quantum Fisher information (QFI) and the Wigner-Yanase skew information (WY). In particular, focusing on the single-qubit state in Eq.~\eqref{eq:000000000001}, the length of the path depicted by the nonunitary evolution of the state $\rho_t$ becomes
\begin{equation}
\label{eq:000000000003}
{\ell_{\gamma}^f}({\rho_0},{\rho_{\tau}}) = \frac{1}{2} \, {\int_0^{\tau}} dt \sqrt{ {h_t^f} } \, \left| \frac{d{\langle{\sigma_x}\rangle_t}}{dt} \right| ~,
\end{equation}
with
\begin{equation}
\label{eq:0000000000004}
{h^{\text{QFI}}_t} := \frac{1 - {\langle{\sigma_z}\rangle_0^2}}{1 - {\langle{\sigma_x}\rangle_t^2} - {\langle{\sigma_z}\rangle_0^2}} ~,
\end{equation}
and
\begin{align}
\label{eq:000000000005}
{h^{\text{WY}}_t} &:= \frac{{\langle{\sigma_x}\rangle_t^2}}{\left({\langle{\sigma_x}\rangle_t^2} + {\langle{\sigma_z}\rangle_0^2}\right) \left(1 - {\langle{\sigma_x}\rangle_t^2} - {\langle{\sigma_z}\rangle_0^2}\right)} \nonumber\\
&+ \frac{2 {\langle{\sigma_z}\rangle_0^2} \left(1 - \sqrt{1 - {\langle{\sigma_x}\rangle_t^2} - {\langle{\sigma_z}\rangle_0^2}} \,\right)}{ {\left({\langle{\sigma_x}\rangle_t^2} + {\langle{\sigma_z}\rangle_0^2} \right)^2} } ~.
\end{align}

Overall, the quantity $(1/2)\sqrt{h_t^f}\left|{d{\langle{\sigma_x}\rangle_t}/dt}\right|$ in Eq.~\eqref{eq:000000000003} signals the speed of evolution respective to the nonunitary dynamics of the single-qubit state, for a given Riemannian metric. Note that this quantity depends on the time-dependent single-qubit observable ${\langle{\sigma_x}\rangle_t}$ that is accessed experimentally. For details in the proof of Eqs.~\eqref{eq:000000000003},~\eqref{eq:0000000000004}, and~\eqref{eq:000000000005}, see the Supplementary Materials~\cite{SupMat}, and also Refs.~\cite{Diego2017,Ingemar_Bengtsson_Zyczkowski,Nielsen_Chuang_infor_geom,MC-paper,Petz-paper} therein. We point out that $\gamma$ is an arbitrary path connecting states $\rho_{0}$ and $\rho_{\tau}$, and its length need not be the shortest one~\cite{PhysRevLett.65.1697}. Indeed, for a given Riemannian metric on the space of quantum states, there exists a geodesic path with minimum length ${\mathcal{L}^f}(\, {\rho_0},{\rho_{\tau}})$ followed by the evolved state $\rho_t$ when going from $\rho_{0}$ to $\rho_{\tau}$. On the one hand, the geodesic length related to the QFI metric is given by the Bures angle, ${\mathcal{L}^{\text{QFI}}}({\rho_0},{\rho_{\tau}}) = \arccos[\sqrt{F({\rho_0},{\rho_{\tau}})}]$, where the Uhlmann fidelity related to initial ${\rho_0}$ and final ${\rho_{\tau}}$ single-qubit states yields
\begin{align}
\label{eq:000000000006}
&F({\rho_0},{\rho_{\tau}}) = \nonumber\\
&\frac{1}{2}\left[1 + {\langle{\sigma_x}\rangle_0}{\langle{\sigma_x}\rangle_{\tau}} + {\langle{\sigma_z}\rangle_0^2} + {\prod_{s = 0,\tau}}\sqrt{ 1 - {\langle{\sigma_x}\rangle_s^2} - {\langle{\sigma_z}\rangle_0^2} } \, \right] ~.
\end{align}

On the other hand, the WY metric implies the geodesic length known as Hellinger angle, ${\mathcal{L}^{\text{WY}}}({\rho_0},{\rho_{\tau}}) = \arccos[A({\rho_0},{\rho_{\tau}})]$, while the quantum affinity for single-qubit states is given by
\begin{align}
\label{eq:000000000007}
&A({\rho_0},{\rho_{\tau}}) = \nonumber\\
&\frac{{\langle{\sigma_x}\rangle_0}{\langle{\sigma_x}\rangle_{\tau}} + {\langle{\sigma_z}\rangle_0^2} + {\prod_{s = 0, \tau}}\left(1 + \sqrt{1 - {\langle{\sigma_x}\rangle_s^2} - {\langle{\sigma_z}\rangle_0^2}} \, \right)}{{\prod_{s = 0, \tau}} \left(\sqrt{1 + \sqrt{ {\langle{\sigma_x}\rangle_s^2} + {\langle{\sigma_z}\rangle_0^2}}} + \sqrt{1 - \sqrt{ {\langle{\sigma_x}\rangle_s^2} + {\langle{\sigma_z}\rangle_0^2} }}  \, \right)} ~.
\end{align}

The geodesic length constitutes a lower bound for the length of the path drawn by the above dynamical evolution, i.e., ${\mathcal{L}^f}({\rho_{0}},{\rho_{\tau}}) \leq {{\ell}^f_{\gamma}}({\rho_0},{\rho_{\tau}})$. Solving this inequality as a function of time, one finds the QSL time related to the nonunitary evolution of the single-qubit state in Eq.~\eqref{eq:000000000001}. In this setting, any distinguishability measure of quantum states gives rise to a different geometric QSL. The contractive Riemannian metric whose geodesic length ${\mathcal{L}^f}$ is effectively tailored to the nonunitary dynamical evolution depicted by the length ${{\ell}^f_{\gamma}}$ is the one that signals the tightest QSL~\cite{Diego2017}. To investigate the tightness of a given geometric QSL, we set the relative deviation
\begin{equation}
\label{eq:000000000008}
{\delta^f_\gamma} := \frac{{{\ell}^f_{\gamma}}({\rho_0},{\rho_{\tau}}) - {\mathcal{L}^f}({\rho_{0}},{\rho_{\tau}})}{{\mathcal{L}^f}({\rho_{0}},{\rho_{\tau}})} ~.
\end{equation}
For a given metric, Eq.~\eqref{eq:000000000008} indicates how far the dynamical evolution is from the respective geodesic path, and is expected to approach zero when both coincide such that the QSL bound saturates. Here, the tightest geometric QSL for the nonunitary dynamics of the single-qubit state is obtained after mi\-ni\-mi\-zing the quantity $\delta^f_\gamma$ over the two aforementioned information-theoretic quantifiers, namely, QFI and WY metrics. In addition, we consider the relative difference $\delta^{\text{QFI}}_\gamma - \delta^{\text{WY}}_\gamma$ as a criterion to testify the tightest QSL, i.e., for ${\delta_{\gamma}^{\text{QFI}}} - {\delta_{\gamma}^{\text{WY}}} > 0$ we have that WY metric assigns the tighter QSL, while for ${\delta_{\gamma}^{\text{QFI}}} - {\delta_{\gamma}^{\text{WY}}} < 0$ the QFI metric sets the tightest lower bound.

In the following, we discuss geometric QSLs for the single-qubit state by focusing on QFI and WY metrics. To do so, we compare numerical simulations of the open-system dynamics with the results calculated using experimental data. The quantum state undergoes a nonunitary evolution exhibiting both non-Markovian and Markovian regimes. In the limit of high concentrations in which the system undergoes Markovian dynamics, we find that the WY metric sets the tighter QSL bound. However, for low concentrations, we observe non-Markovian dynamics, and the relative difference $\delta^{\text{QFI}}_\gamma - \delta^{\text{WY}}_\gamma$ [see Eq.~\eqref{eq:000000000008}] exhibits crossovers between the geometric QSL bounds related to the QFI and WY metrics.

\begin{figure}[!t]
\centering
\includegraphics[scale=1.0]{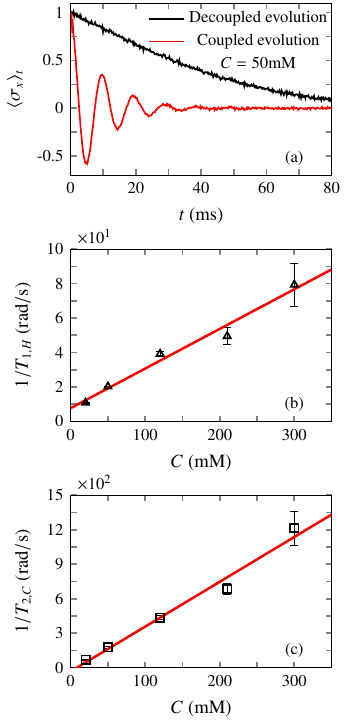}
\caption{(Color online) (a) Coupled and decoupled $^{13}$C FID's for the $50\text{mM}$ concentration, showing the speedup of the decoherence due to scalar relaxation. The deviation from an exponential decay is the result of a small frequency offset. The dependence of ${T_{1,H}}$ (b) and ${T_{2,C}}$ (c) on the Fe(acac)$_3$ concentration, exhibiting the well known linear dependence of the relaxation rate on the concentration of a paramagnetic species~\cite{abragam}.}
\label{FIG01}
\label{fig:FIG01}
\end{figure}



\begin{figure*}[!t]
\begin{center}
\includegraphics[scale=0.85]{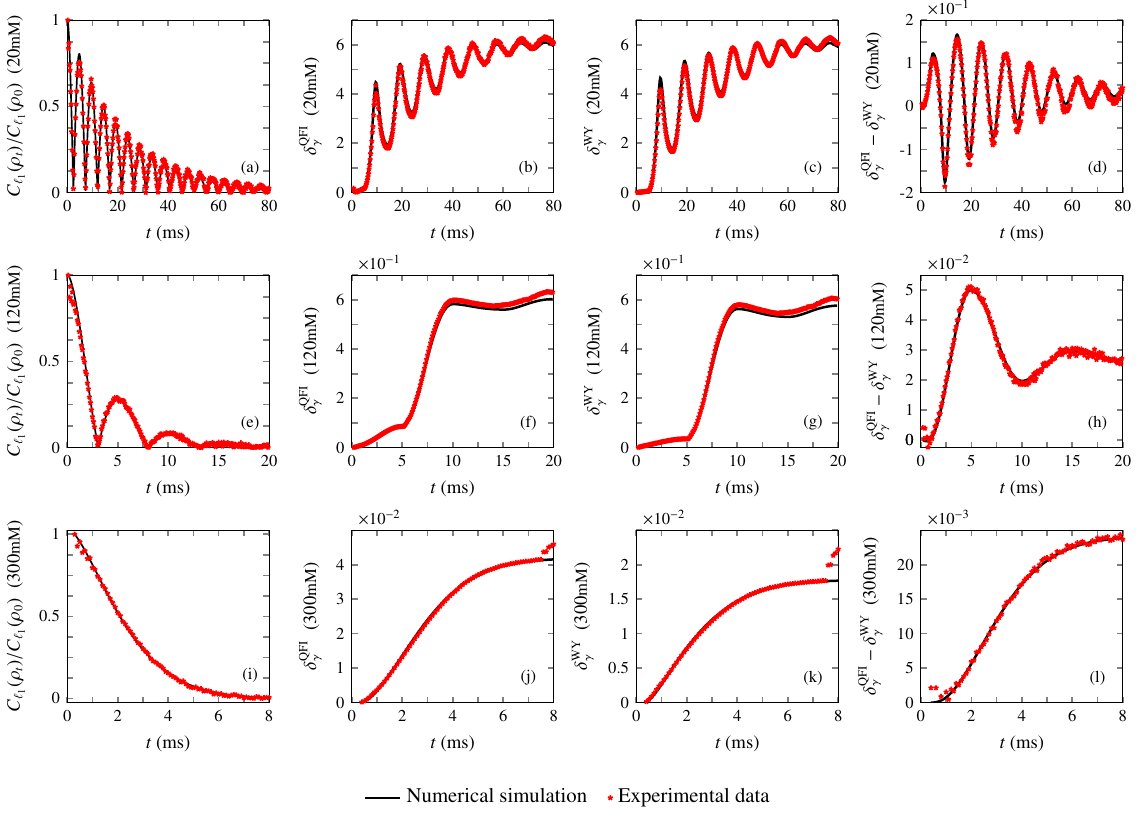}
\caption{(Color online) Plot of the normalized coherence measure ${C_{\ell_1}}({\rho_t})/{C_{\ell_1}}({\rho_0})$ (a,e,i); relative deviations $\delta_\gamma^{\text{QFI,~WY}}$ (b,c,f,g,j,k); and relative difference $\delta_\gamma^{\text{QFI}} - \delta_\gamma^{\text{WY}}$ (d,h,l), for the concentrations $20\text{mM}$, $120\text{mM}$ and $300\text{mM}$. The system is initialized at the single-qubit state ${\rho_0} = (1/2)(\mathbb{I} + {\langle{\sigma_x}\rangle_0}{\sigma_x} + {\langle{\sigma_z}\rangle_0}{\sigma_z})$, with ${\langle{\sigma_x}\rangle_0} = {\langle{\sigma_z}\rangle_0} = 1/\sqrt{2}$. In each panel, the experimental data is shown as red dots, while solid black lines correspond to the results obtained from the numerical simulations, setting the coupling strength $J = 209.1$ Hz, with ${T_{1,H}} = 7.1$ ms and ${T_{2,C}} = 38.55$ ms (a,b,c,d); ${T_{1,H}} = 1.15$ ms and ${T_{2,C}} = 12.8$ ms (e,f,g,h); ${T_{1,H}} = 0.425$ ms and ${T_{2,C}} = 5.49$ ms (i,j,k,l).}
\label{fig:FIG02}
\end{center}
\end{figure*}

\textit{\textbf{Results and Discussion}} --- In Fig.~\ref{fig:FIG01}(a), we show the long decay of the $^{13}$C FID when we decouple the $^{13}$C and $^{1}$H nuclear spins, which makes clear the effect of the $^{1}$H as the main source of decoherence. This effect happens due to the longitudinal relaxation of the $^{1}$H nuclear spin and the scalar $J$ coupling having the same time scale. When ${T_{1,H}}$ $\approx$ $1/J$, the random spin flips on the proton spin caused by the $T_{1,H}$ process make the scalar interaction effectively time-dependent, turning the $^{1}$H into a source of decoherence for the $^{13}$C. Thus, by changing the concentration of Fe(acac)$_3$, we control the correlation time of the system-environment interaction. 

In the presence of paramagnetic ions, the magnetic dipolar interaction of the nuclear spins with the spins of the ion's unpaired electrons is usually much stronger than with other nuclear spins. Thus paramagnetic relaxation becomes the primary nuclear relaxation mechanism, scaling up the inverse of the average distance between the ions and the nucleus, resing in a linear dependence of the relaxation rates, $1/T_{1,H}$ and $1/T_{2,C}$, with the concentration of paramagnetic ions~\cite{abragam}.  Figures~\ref{fig:FIG01}(b) and~\ref{fig:FIG01}(c) show the linear behavior for $1/{T_{1,H}}$ and $1/{T_{2,C}}$, as expected for paramagnetic relaxation.

In Fig.~\ref{fig:FIG02}, we discuss the tightness of the geometric QSL bounds related to QFI and WY metrics, and the Markovian/non-Markovian properties of the nonunitary dynamics of the single-qubit system. The solid lines depict the best fit using the model given in Eq.~\eqref{eq:000000000001}. The system initiates at the state ${\rho_0} = (1/2)(\mathbb{I} + ({\sigma_x} + {\sigma_z})/\sqrt{2})$. The measured values are ${T_{1,H}} = 12 \pm 1$ ms and ${T_{2,C}} = 480\pm 20$ ms for $C = 20\text{mM}$; ${T_{1,H}} = 1.7 \pm 0.2$ ms and ${T_{2,C}} = 87 \pm 3$ ms for $C = 120\text{mM}$; and ${T_{1,H}} = 0.63 \pm 0.08$ ms and ${T_{2,C}} = 29 \pm 2$ ms for $C = 300\text{mM}$. In the numerical simulations, we used the coupling strength $J = 209.1$ Hz, with ${T_{1,H}} = 7.1$ ms and ${T_{2,C}} = 38.55$ ms for $C = 20\text{mM}$ [see Figs.~\ref{fig:FIG02}(a)--\ref{fig:FIG02}(d)]; ${T_{1,H}} = 1.15$ ms and ${T_{2,C}} = 12.8$ ms for $C = 120\text{mM}$ [see Figs.~\ref{fig:FIG02}(e)--\ref{fig:FIG02}(h)]; and ${T_{1,H}} = 0.425$ ms and ${T_{2,C}} = 5.49$ ms for $C = 300\text{mM}$ [see Figs.~\ref{fig:FIG02}(i)--\ref{fig:FIG02}(l)]. The large discrepancy between the measured and simulated values is mainly due to frequency offset errors, which accelerate the signal decay and lead to shorter $T_{1,H}$ and $T_{2,C}$ for fitted values. Such errors come from difficulty in setting the exact resonance frequency when the spectral lines are broad due to the fast relaxation rates used in the experiments. Despite this, note that the measured and simulated values for $T_{1,H}$ have the same order of magnitude. For the physical model in Eq.~\eqref{eq:000000000002}, the relaxation time $T_{1,H}$ is more relevant than $T_{2,C}$. To see this, first we note that ${T_{1,H}}/{T_{2,C}} \ll 1$, i.e., the function $\xi(t)$ [see Eq.~\eqref{eq:000000000002}] will be mainly dominated by $T_{1,H}$ rather than $T_{2,C}$.

In the following, we comment on the non-Markovian and Markovian dynamical signatures of the quantum system. The characterization of non-Markovian dynamics has motivated several efforts to understand the mechanism of information backflow from the environment to the system, and the notion of divisibility of a dynamical map. Historically, the study of information backflow in open quantum systems with the set of measures of non-Markovianity introduced by (i) Breuer, Laine, and Piilo (BLP)~\cite{Breuer2009}; (ii) Rivas, Huelga, and Plenio (RHP)~\cite{PhysRevLett.105.050403}; (iii) Luo, Fu, and Song (LFS)~\cite{PhysRevA.86.044101}. In spite of that, recent studies have addressed the dynamical behavior of quantum coherence measures to characterize non-Markovianity~\cite{PhysRevA.95.042104,PhysRevA.96.022106,SciRep_9_2363_2019}. For a single qubit dissipative channel, it has been shown that the $\ell_1$-norm of coherence exhibits revivals and thus captures non-Markovian signatures of the dynamics in the same way as the BLP measure~\cite{AnnPhys_366_1_2016}. The main idea is that, since the coherence measure monotonically decreases under completely positive and trace preserving (CPTP) incoherent operations, one finds the $\ell_1$-norm of coherence useful for detecting non-Markovianity. For a detailed discussion, see Ref.~\cite{npj_Quantum_Inf_6_55_2020} and references therein. In this setting, we investigate the dynamical behavior of quantum coherence of the two-level system, thus characterizing non-Markovian and Markovian regimes based on the so-called $\ell_1$-norm of coherence. Figures~\ref{fig:FIG02}(a),~\ref{fig:FIG02}(e), and~\ref{fig:FIG02}(i) show the normalized $\ell_1$-norm of coherence of the evolved single-qubit state for different concentrations of Fe(acac)$_3$, with ${C_{\ell_1}}(\rho_t) = {\sum_{j \neq l}} |\langle{j}|{\rho_t}|l\rangle|$~\cite{PhysRevLett.113.140401,RevModPhys.89.041003}. Here, we set $\{|0\rangle,|1\rangle\}$ as the reference basis to evaluate the coherence measure, with ${\sigma_z}|s\rangle = {(-1)^s}|s\rangle~\forall s \in \{0,1\}$. In Fig.~\ref{fig:FIG02}(a), with $C = 20\text{mM}$, the normalized quantum coherence measure exhibits periodic revivals with damped amplitudes, thus vanishing at later times of the dynamics. Figure~\ref{fig:FIG02}(e) shows a qualitatively similar behavior, with the quantum coherence approaching zero faster for $C = 120\text{mM}$. In both cases, the revivals point to the signature of non-Markovian dynamics. In Fig.~\ref{fig:FIG02}(i), the quantum coherence displays a monotonic decay for $C = 300\text{mM}$, a typical behavior of Markovian dynamics~\cite{Breuer2009,AnnPhys_366_1_2016,npj_Quantum_Inf_6_55_2020}. Overall, from low to high concentrations of the paramagnetic salt, we observe the non-Markovian regime for ${T_{1,H}} \approx 1/J$ ($20\text{mM}$ and $120\text{mM}$), while for ${T_{1,H}} \ll$ $1/J$ the dynamics is Markovian ($300\text{mM}$).

Hereafter, we discuss the relative deviation ${\delta_{\gamma}^f}$ [see Eq.~\eqref{eq:000000000008}] for the geometric QSL bound constructed with the QFI metric [Figs.~\ref{fig:FIG02}(b),~\ref{fig:FIG02}(f), and~\ref{fig:FIG02}(j)], and the WY metric [Figs.~\ref{fig:FIG02}(c),~\ref{fig:FIG02}(g), and~\ref{fig:FIG02}(k)]. These quantities are very sensitive to noise since even the small fluctuations observed for short times in the signals of 120 and 300 mM concentrations heavily affect ${\delta_{\gamma}^f}$. This happens due to the time-derivative in Eq.~\eqref{eq:000000000003}. Thus, to avoid undesired numerical errors, we smoothed the data before the numerical integration to evaluate $\ell_{\gamma}^{f}$ for both metrics. The results without smoothing the data are shown in the Supplementary Materials~\cite{SupMat}. Figures~\ref{fig:FIG02}(b) and~\ref{fig:FIG02}(c) show the relative deviations ${\delta_{\gamma}^{\text{QFI}}}$ and ${\delta_{\gamma}^{\text{WY}}}$ within the non-Markovian regime of the dynamics, with $C = 20\text{mM}$. These relative deviations oscillate out of phase with the FID signal. Next, for $C = 120\text{mM}$, Figs.~\ref{fig:FIG02}(f) and~\ref{fig:FIG02}(g) show each relative deviation with a non-monotonic behavior, a fingerprint of non-Markovianity for non-unitary evolution. Figures~\ref{fig:FIG02}(j) and~\ref{fig:FIG02}(k) show that for $C = 300\text{mM}$, both relative deviations behave mo\-no\-to\-ni\-cally as a function of the evolution time of the Markovian dynamics.

To investigate the tightness of the quantum speed limit, we plot in Figs.~\ref{fig:FIG02}(d),~\ref{fig:FIG02}(h), and~\ref{fig:FIG02}(l) the relative difference ${\delta_{\gamma}^{\text{QFI}}} - {\delta_{\gamma}^{\text{WY}}}$ for the concentrations $20\text{mM}$, $120\text{mM}$ and $300\text{mM}$, respectively. Importantly, for ${\delta_{\gamma}^{\text{QFI}}} - {\delta_{\gamma}^{\text{WY}}} > 0$, one finds the tighter speed limit signaled by the WY metric. Otherwise, for ${\delta_{\gamma}^{\text{QFI}}} - {\delta_{\gamma}^{\text{WY}}} < 0$, the QFI metric systematically produces the tightest lower bound to the evolution time. It is worth noting that the tightness of the QSL bound depends on the physical process considered. It turns out that the QSL bound related to the QFI metric is tighter than that speed limit obtained from WY metric for closed quantum systems~\cite{Diego2017}. However, our experiments show the opposite situation for the considered open quantum system dynamics. For $C = 20\text{mM}$, Fig.~\ref{fig:FIG02}(d) shows that the rela\-tive difference oscillates over the evolution time and exhibits crossovers between ${\delta_{\gamma}^{\text{QFI}}}$ and ${\delta_{\gamma}^{\text{WY}}}$, and both QFI and WY metrics give rise to the same geo\-me\-tric QSL whenever ${\delta_{\gamma}^{\text{QFI}}} - {\delta_{\gamma}^{\text{WY}}} = 0$. No crossover occurs after $t \sim 57.4$~ms, and the QSL related to the WY metric turns out to be the tighter one at later times of the dynamics. The tightest geometric QSL is captured by the Wigner-Yanase skew information metric for the two higher concentrations, even for the non-Markovian regime observed for the 120 mM concentration. This is seen in Figs.~\ref{fig:FIG02}(h) and \ref{fig:FIG02}(l).



\textit{\textbf{Conclusions}} --- In summary, we experimentally assess geometric speed limits for the nonunitary dynamics of a qubit encoded on $^{13}$C nuclear spin. We control the relaxation rates of the qubit by adding Fe(acac)$_3$ to the solution. The paramagnetic relaxation makes it possible to vary the relaxation times by two orders of magnitude, and we could observe a transition between non-Markovian to Markovian regimes in the nonunitary reduced dynamics of $^{13}$C spins depending on the salt concentrations. Taking a geometric approach to address the QSL based on the quantum Fisher information (QFI) and Wigner-Yanase (WY) skew information metrics, we find a good agreement between the results of numerical simulations and the experiment. We emphasize the tightness of the QSL depends on the properties of the physical process that is considered, i.e., it is a function of the initial state and the dynamical map that governs the system evolution. For example, for any single-qubit unitary dynamics, it is known that the geometric QSL related to the QFI metric is tighter than the one corresponding to the WY metric. However, one can find instances of nonunitary dynamics of single-qubit states in which the QSL bound constructed with the WY metric is tighter than the QFI one, and vice-versa~\cite{Diego2017}.

The present work is unprecedented in the specialized literature, as it discusses the experimental investigation of QSLs for an open quantum system that is carried out on a controllable Nuclear Magnetic Resonance platform. Past results covered experimental discussion of the QSL derived by Mandelstam-Tamm for the unitary dynamics of nuclear spin systems~\cite{PhysRevA.97.052125}, and also the study of the Mandestam-Tamm and Margolus-Levitin speed limits by using the technique of matter wave interferometry of single atom in an optical trap~\cite{sciadvabj91192021}. Our discussion focuses on geometric QSLs related to the QFI and WY metrics. To the best of our knowledge, QFI and WY metrics cannot be directly measured for general physical processes. It is known that these quantities recover the variance of the dynamical generator for closed systems initialized at pure states~\cite{npjQuantInf_8_56_2022,PhysRevA.107.012414}. However, for nonunitary dynamics and mixed states, bounds on the variance are found that provide, at most, estimates of these quantities~\cite{PhysRevA.73.022324,PhysRevA.87.032324}. To overcome such issues, we recast the geometric QSL bounds in terms of single-qubit observables $\{ {\langle{\sigma_l}\rangle}_{0,t} \}_{l = x,y,z}$ [see Eqs.~\eqref{eq:000000000003}--\eqref{eq:000000000007}], which are probed experimentally.

In our physical system, we find that the tightness of the QSL bound is somehow related to the concentration of the paramagnetic salt that is added to the ensemble of chloroform molecules. By varying such concentration, the strength of the system-reservoir coupling changes and therefore a transition between non-Markovian to Markovian regimes is observed in the single-qubit dynamics encoded in nuclear carbon spins. For lower concentrations ($C = 20$mM), we observe crossovers between the QFI and WY as the tighter one on a non-Markovian regime. In the other two studied cases ($C = 120$mM and $C = 300$mM), the WY metric sets the tighter QSL, both in non-Markovian and Markovian dynamics.

We remark how geometric QSLs are very sensitive to noise, i.e., even tiny fluctuations observed for short times in the data, from low to high concentrations, heavily affect the figure of merit that signals the tighter QSL bound of the two-level system. We note that the relative deviation indicates how much the dynamical evolution differs from the respective geodesic related to the considered metric. The smaller the relative deviation, the tighter the QSL bound for each concentration. We note that the relative deviations ${\delta_{\gamma}^{\text{QFI}}}$ and ${\delta_{\gamma}^{\text{WY}}}$ show similar numerical behaviors for each concentration [see Figs.~\ref{fig:FIG02}(b),~\ref{fig:FIG02}(c),~\ref{fig:FIG02}(f),~\ref{fig:FIG02}(g),~\ref{fig:FIG02}(j), and~\ref{fig:FIG02}(l)]. This means that, for the physical process considered, the QFI and WY metrics provide QSL bounds that are close to each other. Finally, this result might suggest that neither QFI or WY metrics can be the fundamental one, which can foster the investigation of new proposals of {\it bona fide} QSL metrics.


\textit{\textbf{Acknowledgments}} --- This work was supported by the Brazilian ministries MEC and MCTIC, and the Brazilian funding agencies CNPq (Grant No. 304891/2022-3), FAPESP (Grant No. 2017/03727-0), Coordena\c{c}\~{a}o de Aperfei\c{c}oamento de Pessoal de N\'{i}vel Superior--Brasil (CAPES) (Finance Code 001), and the Brazilian National Institute of Science and Technology of Quantum Information (INCT-IQ). D. P. P. also acknowledges Funda\c{c}\~{a}o de Amparo \`{a} Pesquisa e ao Desenvolvimento Cient\'{i}fico e Tecnol\'{o}gico do Maranh\~{a}o (FAPEMA).



\begin{thebibliography}{75}%
\makeatletter
\providecommand \@ifxundefined [1]{%
 \@ifx{#1\undefined}
}%
\providecommand \@ifnum [1]{%
 \ifnum #1\expandafter \@firstoftwo
 \else \expandafter \@secondoftwo
 \fi
}%
\providecommand \@ifx [1]{%
 \ifx #1\expandafter \@firstoftwo
 \else \expandafter \@secondoftwo
 \fi
}%
\providecommand \natexlab [1]{#1}%
\providecommand \enquote  [1]{``#1''}%
\providecommand \bibnamefont  [1]{#1}%
\providecommand \bibfnamefont [1]{#1}%
\providecommand \citenamefont [1]{#1}%
\providecommand \href@noop [0]{\@secondoftwo}%
\providecommand \href [0]{\begingroup \@sanitize@url \@href}%
\providecommand \@href[1]{\@@startlink{#1}\@@href}%
\providecommand \@@href[1]{\endgroup#1\@@endlink}%
\providecommand \@sanitize@url [0]{\catcode `\\12\catcode `\$12\catcode
  `\&12\catcode `\#12\catcode `\^12\catcode `\_12\catcode `\%12\relax}%
\providecommand \@@startlink[1]{}%
\providecommand \@@endlink[0]{}%
\providecommand \url  [0]{\begingroup\@sanitize@url \@url }%
\providecommand \@url [1]{\endgroup\@href {#1}{\urlprefix }}%
\providecommand \urlprefix  [0]{URL }%
\providecommand \Eprint [0]{\href }%
\providecommand \doibase [0]{http://dx.doi.org/}%
\providecommand \selectlanguage [0]{\@gobble}%
\providecommand \bibinfo  [0]{\@secondoftwo}%
\providecommand \bibfield  [0]{\@secondoftwo}%
\providecommand \translation [1]{[#1]}%
\providecommand \BibitemOpen [0]{}%
\providecommand \bibitemStop [0]{}%
\providecommand \bibitemNoStop [0]{.\EOS\space}%
\providecommand \EOS [0]{\spacefactor3000\relax}%
\providecommand \BibitemShut  [1]{\csname bibitem#1\endcsname}%
\let\auto@bib@innerbib\@empty
\bibitem [{\citenamefont {Aharonov}\ and\ \citenamefont
  {Bohm}(1961)}]{PhysRev.122.1649}%
  \BibitemOpen
  \bibfield  {author} {\bibinfo {author} {\bibfnamefont {Y.}~\bibnamefont
  {Aharonov}}\ and\ \bibinfo {author} {\bibfnamefont {D.}~\bibnamefont
  {Bohm}},\ }\bibfield  {title} {\enquote {\bibinfo {title} {Time in the
  {Q}uantum {T}heory and the {U}ncertainty {R}elation for {T}ime and
  {E}nergy},}\ }\href {\doibase 10.1103/PhysRev.122.1649} {\bibfield  {journal}
  {\bibinfo  {journal} {Phys. Rev.}\ }\textbf {\bibinfo {volume} {122}},\
  \bibinfo {pages} {1649} (\bibinfo {year} {1961})}\BibitemShut {NoStop}%
\bibitem [{\citenamefont {Pfeifer}\ and\ \citenamefont
  {Fr\"ohlich}(1995)}]{RevModPhys.67.759}%
  \BibitemOpen
  \bibfield  {author} {\bibinfo {author} {\bibfnamefont {P.}~\bibnamefont
  {Pfeifer}}\ and\ \bibinfo {author} {\bibfnamefont {J.}~\bibnamefont
  {Fr\"ohlich}},\ }\bibfield  {title} {\enquote {\bibinfo {title} {Generalized
  time-energy uncertainty relations and bounds on lifetimes of resonances},}\
  }\href {\doibase 10.1103/RevModPhys.67.759} {\bibfield  {journal} {\bibinfo
  {journal} {Rev. Mod. Phys.}\ }\textbf {\bibinfo {volume} {67}},\ \bibinfo
  {pages} {759} (\bibinfo {year} {1995})}\BibitemShut {NoStop}%
\bibitem [{\citenamefont {Mandelstam}\ and\ \citenamefont
  {Tamm}(1991)}]{Mandelstam}%
  \BibitemOpen
  \bibfield  {author} {\bibinfo {author} {\bibfnamefont {L.}~\bibnamefont
  {Mandelstam}}\ and\ \bibinfo {author} {\bibfnamefont {I.}~\bibnamefont
  {Tamm}},\ }\enquote {\bibinfo {title} {The uncertainty relation between
  energy and time in non-relativistic quantum mechanics},}\ in\ \href {\doibase
  10.1007/978-3-642-74626-0_8} {\emph {\bibinfo {booktitle} {Selected
  {P}apers}}},\ \bibinfo {editor} {edited by\ \bibinfo {editor} {\bibfnamefont
  {B.~M.}\ \bibnamefont {Bolotovskii}}, \bibinfo {editor} {\bibfnamefont
  {V.~Y.}\ \bibnamefont {Frenkel}}, \ and\ \bibinfo {editor} {\bibfnamefont
  {R.}~\bibnamefont {Peierls}}}\ (\bibinfo  {publisher} {Springer},\ \bibinfo
  {address} {Berlin, {H}eidelberg},\ \bibinfo {year} {1991})\ pp.\ \bibinfo
  {pages} {115--123}\BibitemShut {NoStop}%
\bibitem [{\citenamefont {Margolus}\ and\ \citenamefont
  {Levitin}(1998)}]{1992_PhysicaD_120_188}%
  \BibitemOpen
  \bibfield  {author} {\bibinfo {author} {\bibfnamefont {N.}~\bibnamefont
  {Margolus}}\ and\ \bibinfo {author} {\bibfnamefont {L.~B.}\ \bibnamefont
  {Levitin}},\ }\bibfield  {title} {\enquote {\bibinfo {title} {The maximum
  speed of dynamical evolution},}\ }\href {\doibase
  10.1016/S0167-2789(98)00054-2} {\bibfield  {journal} {\bibinfo  {journal}
  {Physica D}\ }\textbf {\bibinfo {volume} {120}},\ \bibinfo {pages} {188}
  (\bibinfo {year} {1998})}\BibitemShut {NoStop}%
\bibitem [{\citenamefont {Taddei}\ \emph {et~al.}(2013)\citenamefont {Taddei},
  \citenamefont {Escher}, \citenamefont {Davidovich},\ and\ \citenamefont
  {de~Matos~Filho}}]{Taddei2013}%
  \BibitemOpen
  \bibfield  {author} {\bibinfo {author} {\bibfnamefont {M.~M.}\ \bibnamefont
  {Taddei}}, \bibinfo {author} {\bibfnamefont {B.~M.}\ \bibnamefont {Escher}},
  \bibinfo {author} {\bibfnamefont {L.}~\bibnamefont {Davidovich}}, \ and\
  \bibinfo {author} {\bibfnamefont {R.~L.}\ \bibnamefont {de~Matos~Filho}},\
  }\bibfield  {title} {\enquote {\bibinfo {title} {Quantum speed limit for
  physical processes},}\ }\href {\doibase 10.1103/PhysRevLett.110.050402}
  {\bibfield  {journal} {\bibinfo  {journal} {Phys. Rev. Lett.}\ }\textbf
  {\bibinfo {volume} {110}},\ \bibinfo {pages} {050402} (\bibinfo {year}
  {2013})}\BibitemShut {NoStop}%
\bibitem [{\citenamefont {del Campo}\ \emph {et~al.}(2013)\citenamefont {del
  Campo}, \citenamefont {Egusquiza}, \citenamefont {Plenio},\ and\
  \citenamefont {Huelga}}]{delcampo2013}%
  \BibitemOpen
  \bibfield  {author} {\bibinfo {author} {\bibfnamefont {A.}~\bibnamefont {del
  Campo}}, \bibinfo {author} {\bibfnamefont {I.~L.}\ \bibnamefont {Egusquiza}},
  \bibinfo {author} {\bibfnamefont {M.~B.}\ \bibnamefont {Plenio}}, \ and\
  \bibinfo {author} {\bibfnamefont {S.~F.}\ \bibnamefont {Huelga}},\ }\bibfield
   {title} {\enquote {\bibinfo {title} {Quantum speed limits in open system
  dynamics},}\ }\href {\doibase 10.1103/PhysRevLett.110.050403} {\bibfield
  {journal} {\bibinfo  {journal} {Phys. Rev. Lett.}\ }\textbf {\bibinfo
  {volume} {110}},\ \bibinfo {pages} {050403} (\bibinfo {year}
  {2013})}\BibitemShut {NoStop}%
\bibitem [{\citenamefont {Deffner}\ and\ \citenamefont
  {Lutz}(2013)}]{PhysRevLett.111.010402}%
  \BibitemOpen
  \bibfield  {author} {\bibinfo {author} {\bibfnamefont {S.}~\bibnamefont
  {Deffner}}\ and\ \bibinfo {author} {\bibfnamefont {E.}~\bibnamefont {Lutz}},\
  }\bibfield  {title} {\enquote {\bibinfo {title} {Quantum {S}peed {L}imit for
  {N}on-{M}arkovian {D}ynamics},}\ }\href {\doibase
  10.1103/PhysRevLett.111.010402} {\bibfield  {journal} {\bibinfo  {journal}
  {Phys. Rev. Lett.}\ }\textbf {\bibinfo {volume} {111}},\ \bibinfo {pages}
  {010402} (\bibinfo {year} {2013})}\BibitemShut {NoStop}%
\bibitem [{\citenamefont {Okuyama}\ and\ \citenamefont
  {Ohzeki}(2018)}]{PhysRevLett.120.070402}%
  \BibitemOpen
  \bibfield  {author} {\bibinfo {author} {\bibfnamefont {M.}~\bibnamefont
  {Okuyama}}\ and\ \bibinfo {author} {\bibfnamefont {M.}~\bibnamefont
  {Ohzeki}},\ }\bibfield  {title} {\enquote {\bibinfo {title} {Quantum {S}peed
  {L}imit is {N}ot {Q}uantum},}\ }\href {\doibase
  10.1103/PhysRevLett.120.070402} {\bibfield  {journal} {\bibinfo  {journal}
  {Phys. Rev. Lett.}\ }\textbf {\bibinfo {volume} {120}},\ \bibinfo {pages}
  {070402} (\bibinfo {year} {2018})}\BibitemShut {NoStop}%
\bibitem [{\citenamefont {Campaioli}\ \emph {et~al.}(2019)\citenamefont
  {Campaioli}, \citenamefont {Pollock},\ and\ \citenamefont
  {Modi}}]{Quantum_3_168_2019}%
  \BibitemOpen
  \bibfield  {author} {\bibinfo {author} {\bibfnamefont {F.}~\bibnamefont
  {Campaioli}}, \bibinfo {author} {\bibfnamefont {F.~A.}\ \bibnamefont
  {Pollock}}, \ and\ \bibinfo {author} {\bibfnamefont {K.}~\bibnamefont
  {Modi}},\ }\bibfield  {title} {\enquote {\bibinfo {title} {Tight, robust, and
  feasible quantum speed limits for open dynamics},}\ }\href {\doibase
  10.22331/q-2019-08-05-168} {\bibfield  {journal} {\bibinfo  {journal}
  {Quantum}\ }\textbf {\bibinfo {volume} {3}},\ \bibinfo {pages} {168}
  (\bibinfo {year} {2019})}\BibitemShut {NoStop}%
\bibitem [{\citenamefont {O'Connor}\ \emph {et~al.}(2021)\citenamefont
  {O'Connor}, \citenamefont {Guarnieri},\ and\ \citenamefont
  {Campbell}}]{PhysRevA.103.022210}%
  \BibitemOpen
  \bibfield  {author} {\bibinfo {author} {\bibfnamefont {E.}~\bibnamefont
  {O'Connor}}, \bibinfo {author} {\bibfnamefont {G.}~\bibnamefont {Guarnieri}},
  \ and\ \bibinfo {author} {\bibfnamefont {S.}~\bibnamefont {Campbell}},\
  }\bibfield  {title} {\enquote {\bibinfo {title} {Action quantum speed
  limits},}\ }\href {\doibase 10.1103/PhysRevA.103.022210} {\bibfield
  {journal} {\bibinfo  {journal} {Phys. Rev. A}\ }\textbf {\bibinfo {volume}
  {103}},\ \bibinfo {pages} {022210} (\bibinfo {year} {2021})}\BibitemShut
  {NoStop}%
\bibitem [{\citenamefont {Lokutsievskiy}\ and\ \citenamefont
  {Pechen}(2021)}]{JPhysAMathTheor_54_395304_2021}%
  \BibitemOpen
  \bibfield  {author} {\bibinfo {author} {\bibfnamefont {L.}~\bibnamefont
  {Lokutsievskiy}}\ and\ \bibinfo {author} {\bibfnamefont {A.}~\bibnamefont
  {Pechen}},\ }\bibfield  {title} {\enquote {\bibinfo {title} {Reachable sets
  for two-level open quantum systems driven by coherent and incoherent
  controls},}\ }\href {\doibase 10.1088/1751-8121/ac19f8} {\bibfield  {journal}
  {\bibinfo  {journal} {J. Phys. A: Math. Theor.}\ }\textbf {\bibinfo {volume}
  {54}},\ \bibinfo {pages} {395304} (\bibinfo {year} {2021})}\BibitemShut
  {NoStop}%
\bibitem [{\citenamefont {Impens}\ \emph {et~al.}(2021)\citenamefont {Impens},
  \citenamefont {D'Angelis}, \citenamefont {Pinheiro},\ and\ \citenamefont
  {Gu\'ery-Odelin}}]{PhysRevA.104.052620}%
  \BibitemOpen
  \bibfield  {author} {\bibinfo {author} {\bibfnamefont {F.}~\bibnamefont
  {Impens}}, \bibinfo {author} {\bibfnamefont {F.~M.}\ \bibnamefont
  {D'Angelis}}, \bibinfo {author} {\bibfnamefont {F.~A.}\ \bibnamefont
  {Pinheiro}}, \ and\ \bibinfo {author} {\bibfnamefont {D.}~\bibnamefont
  {Gu\'ery-Odelin}},\ }\bibfield  {title} {\enquote {\bibinfo {title} {Time
  scaling and quantum speed limit in non-{H}ermitian {H}amiltonians},}\ }\href
  {\doibase 10.1103/PhysRevA.104.052620} {\bibfield  {journal} {\bibinfo
  {journal} {Phys. Rev. A}\ }\textbf {\bibinfo {volume} {104}},\ \bibinfo
  {pages} {052620} (\bibinfo {year} {2021})}\BibitemShut {NoStop}%
\bibitem [{\citenamefont {Mohan}\ \emph {et~al.}(2022)\citenamefont {Mohan},
  \citenamefont {Das},\ and\ \citenamefont {Pati}}]{arXiv:2110.13193}%
  \BibitemOpen
  \bibfield  {author} {\bibinfo {author} {\bibfnamefont {B.}~\bibnamefont
  {Mohan}}, \bibinfo {author} {\bibfnamefont {S.}~\bibnamefont {Das}}, \ and\
  \bibinfo {author} {\bibfnamefont {A.~K.}\ \bibnamefont {Pati}},\ }\bibfield
  {title} {\enquote {\bibinfo {title} {Quantum speed limits for information and
  coherence},}\ }\href {\doibase 10.1088/1367-2630/ac753c} {\bibfield
  {journal} {\bibinfo  {journal} {New J. Phys.}\ }\textbf {\bibinfo {volume}
  {24}},\ \bibinfo {pages} {065003} (\bibinfo {year} {2022})}\BibitemShut
  {NoStop}%
\bibitem [{\citenamefont {Campaioli}\ \emph {et~al.}(2022)\citenamefont
  {Campaioli}, \citenamefont {Yu}, \citenamefont {Pollock},\ and\ \citenamefont
  {Modi}}]{arXiv:2004.03078}%
  \BibitemOpen
  \bibfield  {author} {\bibinfo {author} {\bibfnamefont {F.}~\bibnamefont
  {Campaioli}}, \bibinfo {author} {\bibfnamefont {C.-S.}\ \bibnamefont {Yu}},
  \bibinfo {author} {\bibfnamefont {F.~A.}\ \bibnamefont {Pollock}}, \ and\
  \bibinfo {author} {\bibfnamefont {K.}~\bibnamefont {Modi}},\ }\bibfield
  {title} {\enquote {\bibinfo {title} {Resource speed limits: {M}aximal rate of
  resource variation},}\ }\href {\doibase 10.1088/1367-2630/ac7346} {\bibfield
  {journal} {\bibinfo  {journal} {New J. Phys.}\ }\textbf {\bibinfo {volume}
  {24}},\ \bibinfo {pages} {065001} (\bibinfo {year} {2022})}\BibitemShut
  {NoStop}%
\bibitem [{\citenamefont {Pires}(2022)}]{PhysRevA.106.012403}%
  \BibitemOpen
  \bibfield  {author} {\bibinfo {author} {\bibfnamefont {D.~P.}\ \bibnamefont
  {Pires}},\ }\bibfield  {title} {\enquote {\bibinfo {title} {Unified entropies
  and quantum speed limits for nonunitary dynamics},}\ }\href {\doibase
  10.1103/PhysRevA.106.012403} {\bibfield  {journal} {\bibinfo  {journal}
  {Phys. Rev. A}\ }\textbf {\bibinfo {volume} {106}},\ \bibinfo {pages}
  {012403} (\bibinfo {year} {2022})}\BibitemShut {NoStop}%
\bibitem [{\citenamefont {Ness}\ \emph {et~al.}(2022)\citenamefont {Ness},
  \citenamefont {Alberti},\ and\ \citenamefont
  {Sagi}}]{PhysRevLett.129.140403}%
  \BibitemOpen
  \bibfield  {author} {\bibinfo {author} {\bibfnamefont {G.}~\bibnamefont
  {Ness}}, \bibinfo {author} {\bibfnamefont {A.}~\bibnamefont {Alberti}}, \
  and\ \bibinfo {author} {\bibfnamefont {Y.}~\bibnamefont {Sagi}},\ }\bibfield
  {title} {\enquote {\bibinfo {title} {Quantum {S}peed {L}imit for {S}tates
  with a {B}ounded {E}nergy {S}pectrum},}\ }\href {\doibase
  10.1103/PhysRevLett.129.140403} {\bibfield  {journal} {\bibinfo  {journal}
  {Phys. Rev. Lett.}\ }\textbf {\bibinfo {volume} {129}},\ \bibinfo {pages}
  {140403} (\bibinfo {year} {2022})}\BibitemShut {NoStop}%
\bibitem [{\citenamefont {Fogarty}\ \emph {et~al.}(2020)\citenamefont
  {Fogarty}, \citenamefont {Deffner}, \citenamefont {Busch},\ and\
  \citenamefont {Campbell}}]{PhysRevLett.124.110601}%
  \BibitemOpen
  \bibfield  {author} {\bibinfo {author} {\bibfnamefont {T.}~\bibnamefont
  {Fogarty}}, \bibinfo {author} {\bibfnamefont {S.}~\bibnamefont {Deffner}},
  \bibinfo {author} {\bibfnamefont {T.}~\bibnamefont {Busch}}, \ and\ \bibinfo
  {author} {\bibfnamefont {S.}~\bibnamefont {Campbell}},\ }\bibfield  {title}
  {\enquote {\bibinfo {title} {Orthogonality {C}atastrophe as a {C}onsequence
  of the {Q}uantum {S}peed {L}imit},}\ }\href {\doibase
  10.1103/PhysRevLett.124.110601} {\bibfield  {journal} {\bibinfo  {journal}
  {Phys. Rev. Lett.}\ }\textbf {\bibinfo {volume} {124}},\ \bibinfo {pages}
  {110601} (\bibinfo {year} {2020})}\BibitemShut {NoStop}%
\bibitem [{\citenamefont {Shao}\ \emph {et~al.}(2020)\citenamefont {Shao},
  \citenamefont {Liu}, \citenamefont {Zhang}, \citenamefont {Yuan},\ and\
  \citenamefont {Liu}}]{PhysRevResearch.2.023299}%
  \BibitemOpen
  \bibfield  {author} {\bibinfo {author} {\bibfnamefont {Y.}~\bibnamefont
  {Shao}}, \bibinfo {author} {\bibfnamefont {B.}~\bibnamefont {Liu}}, \bibinfo
  {author} {\bibfnamefont {M.}~\bibnamefont {Zhang}}, \bibinfo {author}
  {\bibfnamefont {H.}~\bibnamefont {Yuan}}, \ and\ \bibinfo {author}
  {\bibfnamefont {J.}~\bibnamefont {Liu}},\ }\bibfield  {title} {\enquote
  {\bibinfo {title} {Operational definition of a quantum speed limit},}\ }\href
  {\doibase 10.1103/PhysRevResearch.2.023299} {\bibfield  {journal} {\bibinfo
  {journal} {Phys. Rev. Research}\ }\textbf {\bibinfo {volume} {2}},\ \bibinfo
  {pages} {023299} (\bibinfo {year} {2020})}\BibitemShut {NoStop}%
\bibitem [{\citenamefont {Puebla}\ \emph {et~al.}(2020)\citenamefont {Puebla},
  \citenamefont {Deffner},\ and\ \citenamefont
  {Campbell}}]{PhysRevResearch.2.032020}%
  \BibitemOpen
  \bibfield  {author} {\bibinfo {author} {\bibfnamefont {R.}~\bibnamefont
  {Puebla}}, \bibinfo {author} {\bibfnamefont {S.}~\bibnamefont {Deffner}}, \
  and\ \bibinfo {author} {\bibfnamefont {S.}~\bibnamefont {Campbell}},\
  }\bibfield  {title} {\enquote {\bibinfo {title} {Kibble-{Z}urek scaling in
  quantum speed limits for shortcuts to adiabaticity},}\ }\href {\doibase
  10.1103/PhysRevResearch.2.032020} {\bibfield  {journal} {\bibinfo  {journal}
  {Phys. Rev. Research}\ }\textbf {\bibinfo {volume} {2}},\ \bibinfo {pages}
  {032020} (\bibinfo {year} {2020})}\BibitemShut {NoStop}%
\bibitem [{\citenamefont {Kobayashi}\ and\ \citenamefont
  {Yamamoto}(2020)}]{PhysRevA.102.042606}%
  \BibitemOpen
  \bibfield  {author} {\bibinfo {author} {\bibfnamefont {K.}~\bibnamefont
  {Kobayashi}}\ and\ \bibinfo {author} {\bibfnamefont {N.}~\bibnamefont
  {Yamamoto}},\ }\bibfield  {title} {\enquote {\bibinfo {title} {Quantum speed
  limit for robust state characterization and engineering},}\ }\href {\doibase
  10.1103/PhysRevA.102.042606} {\bibfield  {journal} {\bibinfo  {journal}
  {Phys. Rev. A}\ }\textbf {\bibinfo {volume} {102}},\ \bibinfo {pages}
  {042606} (\bibinfo {year} {2020})}\BibitemShut {NoStop}%
\bibitem [{\citenamefont {Lam}\ \emph {et~al.}(2021)\citenamefont {Lam},
  \citenamefont {Peter}, \citenamefont {Groh}, \citenamefont {Alt},
  \citenamefont {Robens}, \citenamefont {Meschede}, \citenamefont {Negretti},
  \citenamefont {Montangero}, \citenamefont {Calarco},\ and\ \citenamefont
  {Alberti}}]{PhysRevX.11.011035}%
  \BibitemOpen
  \bibfield  {author} {\bibinfo {author} {\bibfnamefont {M.~R.}\ \bibnamefont
  {Lam}}, \bibinfo {author} {\bibfnamefont {N.}~\bibnamefont {Peter}}, \bibinfo
  {author} {\bibfnamefont {T.}~\bibnamefont {Groh}}, \bibinfo {author}
  {\bibfnamefont {W.}~\bibnamefont {Alt}}, \bibinfo {author} {\bibfnamefont
  {C.}~\bibnamefont {Robens}}, \bibinfo {author} {\bibfnamefont
  {D.}~\bibnamefont {Meschede}}, \bibinfo {author} {\bibfnamefont
  {A.}~\bibnamefont {Negretti}}, \bibinfo {author} {\bibfnamefont
  {S.}~\bibnamefont {Montangero}}, \bibinfo {author} {\bibfnamefont
  {T.}~\bibnamefont {Calarco}}, \ and\ \bibinfo {author} {\bibfnamefont
  {A.}~\bibnamefont {Alberti}},\ }\bibfield  {title} {\enquote {\bibinfo
  {title} {Demonstration of {Q}uantum {B}rachistochrones between {D}istant
  {S}tates of an {A}tom},}\ }\href {\doibase 10.1103/PhysRevX.11.011035}
  {\bibfield  {journal} {\bibinfo  {journal} {Phys. Rev. X}\ }\textbf {\bibinfo
  {volume} {11}},\ \bibinfo {pages} {011035} (\bibinfo {year}
  {2021})}\BibitemShut {NoStop}%
\bibitem [{\citenamefont {del Campo}(2021)}]{PhysRevLett.126.180603}%
  \BibitemOpen
  \bibfield  {author} {\bibinfo {author} {\bibfnamefont {A.}~\bibnamefont {del
  Campo}},\ }\bibfield  {title} {\enquote {\bibinfo {title} {Probing {Q}uantum
  {S}peed {L}imits with {U}ltracold {G}ases},}\ }\href {\doibase
  10.1103/PhysRevLett.126.180603} {\bibfield  {journal} {\bibinfo  {journal}
  {Phys. Rev. Lett.}\ }\textbf {\bibinfo {volume} {126}},\ \bibinfo {pages}
  {180603} (\bibinfo {year} {2021})}\BibitemShut {NoStop}%
\bibitem [{\citenamefont {Mohan}\ and\ \citenamefont
  {Pati}(2021)}]{arXiv:2006.14523}%
  \BibitemOpen
  \bibfield  {author} {\bibinfo {author} {\bibfnamefont {B.}~\bibnamefont
  {Mohan}}\ and\ \bibinfo {author} {\bibfnamefont {A.~K.}\ \bibnamefont
  {Pati}},\ }\bibfield  {title} {\enquote {\bibinfo {title} {Reverse quantum
  speed limit: {H}ow slowly a quantum battery can discharge},}\ }\href
  {\doibase 10.1103/PhysRevA.104.042209} {\bibfield  {journal} {\bibinfo
  {journal} {Phys. Rev. A}\ }\textbf {\bibinfo {volume} {104}},\ \bibinfo
  {pages} {042209} (\bibinfo {year} {2021})}\BibitemShut {NoStop}%
\bibitem [{\citenamefont {Pires}\ \emph {et~al.}(2021)\citenamefont {Pires},
  \citenamefont {Modi},\ and\ \citenamefont {C\'eleri}}]{PhysRevE.103.032105}%
  \BibitemOpen
  \bibfield  {author} {\bibinfo {author} {\bibfnamefont {D.~P.}\ \bibnamefont
  {Pires}}, \bibinfo {author} {\bibfnamefont {K.}~\bibnamefont {Modi}}, \ and\
  \bibinfo {author} {\bibfnamefont {L.~C.}\ \bibnamefont {C\'eleri}},\
  }\bibfield  {title} {\enquote {\bibinfo {title} {Bounding generalized
  relative entropies: {N}onasymptotic quantum speed limits},}\ }\href {\doibase
  10.1103/PhysRevE.103.032105} {\bibfield  {journal} {\bibinfo  {journal}
  {Phys. Rev. E}\ }\textbf {\bibinfo {volume} {103}},\ \bibinfo {pages}
  {032105} (\bibinfo {year} {2021})}\BibitemShut {NoStop}%
\bibitem [{\citenamefont {Pires}\ and\ \citenamefont
  {de~Oliveira}(2021)}]{PhysRevA.104.052223}%
  \BibitemOpen
  \bibfield  {author} {\bibinfo {author} {\bibfnamefont {D.~P.}\ \bibnamefont
  {Pires}}\ and\ \bibinfo {author} {\bibfnamefont {T.~R.}\ \bibnamefont
  {de~Oliveira}},\ }\bibfield  {title} {\enquote {\bibinfo {title} {Relative
  purity, speed of fluctuations, and bounds on equilibration times},}\ }\href
  {\doibase 10.1103/PhysRevA.104.052223} {\bibfield  {journal} {\bibinfo
  {journal} {Phys. Rev. A}\ }\textbf {\bibinfo {volume} {104}},\ \bibinfo
  {pages} {052223} (\bibinfo {year} {2021})}\BibitemShut {NoStop}%
\bibitem [{\citenamefont {Aghion}\ and\ \citenamefont
  {Green}(2023)}]{arXiv:2204.10368}%
  \BibitemOpen
  \bibfield  {author} {\bibinfo {author} {\bibfnamefont {E.}~\bibnamefont
  {Aghion}}\ and\ \bibinfo {author} {\bibfnamefont {J.~R.}\ \bibnamefont
  {Green}},\ }\bibfield  {title} {\enquote {\bibinfo {title} {Thermodynamic
  speed limits for mechanical work},}\ }\href {\doibase
  10.1088/1751-8121/acb5d6} {\bibfield  {journal} {\bibinfo  {journal} {J.
  Phys. A: Math. Theor.}\ }\textbf {\bibinfo {volume} {56}},\ \bibinfo {pages}
  {05LT01} (\bibinfo {year} {2023})}\BibitemShut {NoStop}%
\bibitem [{\citenamefont {Hasegawa}(2023)}]{arXiv:2203.12421}%
  \BibitemOpen
  \bibfield  {author} {\bibinfo {author} {\bibfnamefont {Y.}~\bibnamefont
  {Hasegawa}},\ }\bibfield  {title} {\enquote {\bibinfo {title} {Unifying speed
  limit, thermodynamic uncertainty relation and {H}eisenberg principle via
  bulk-boundary correspondence},}\ }\href {\doibase 10.1038/s41467-023-38074-8}
  {\bibfield  {journal} {\bibinfo  {journal} {Nat. Commun.}\ }\textbf {\bibinfo
  {volume} {14}},\ \bibinfo {pages} {2828} (\bibinfo {year}
  {2023})}\BibitemShut {NoStop}%
\bibitem [{\citenamefont {Gong}\ and\ \citenamefont
  {Hamazaki}(2022)}]{IntJModPhysB_36_2230007_2022}%
  \BibitemOpen
  \bibfield  {author} {\bibinfo {author} {\bibfnamefont {Z.}~\bibnamefont
  {Gong}}\ and\ \bibinfo {author} {\bibfnamefont {R.}~\bibnamefont
  {Hamazaki}},\ }\bibfield  {title} {\enquote {\bibinfo {title} {Bounds in
  nonequilibrium quantum dynamics},}\ }\href {\doibase
  10.1142/S0217979222300079} {\bibfield  {journal} {\bibinfo  {journal} {Int.
  J. Mod. Phys. B}\ }\textbf {\bibinfo {volume} {36}},\ \bibinfo {pages}
  {2230007} (\bibinfo {year} {2022})}\BibitemShut {NoStop}%
\bibitem [{\citenamefont {Pires}\ \emph {et~al.}(2016)\citenamefont {Pires},
  \citenamefont {Cianciaruso}, \citenamefont {C\'eleri}, \citenamefont
  {Adesso},\ and\ \citenamefont {Soares-Pinto}}]{Diego2017}%
  \BibitemOpen
  \bibfield  {author} {\bibinfo {author} {\bibfnamefont {D.~P.}\ \bibnamefont
  {Pires}}, \bibinfo {author} {\bibfnamefont {M.}~\bibnamefont {Cianciaruso}},
  \bibinfo {author} {\bibfnamefont {L.~C.}\ \bibnamefont {C\'eleri}}, \bibinfo
  {author} {\bibfnamefont {G.}~\bibnamefont {Adesso}}, \ and\ \bibinfo {author}
  {\bibfnamefont {D.~O.}\ \bibnamefont {Soares-Pinto}},\ }\bibfield  {title}
  {\enquote {\bibinfo {title} {Generalized {G}eometric {Q}uantum {S}peed
  {L}imits},}\ }\href {\doibase 10.1103/PhysRevX.6.021031} {\bibfield
  {journal} {\bibinfo  {journal} {Phys. Rev. X}\ }\textbf {\bibinfo {volume}
  {6}},\ \bibinfo {pages} {021031} (\bibinfo {year} {2016})}\BibitemShut
  {NoStop}%
\bibitem [{\citenamefont {Deffner}\ and\ \citenamefont
  {Campbell}(2017)}]{Deffner_2017}%
  \BibitemOpen
  \bibfield  {author} {\bibinfo {author} {\bibfnamefont {S.}~\bibnamefont
  {Deffner}}\ and\ \bibinfo {author} {\bibfnamefont {S.}~\bibnamefont
  {Campbell}},\ }\bibfield  {title} {\enquote {\bibinfo {title} {Quantum speed
  limits: from {H}eisenberg's uncertainty principle to optimal quantum
  control},}\ }\href {\doibase 10.1088/1751-8121/aa86c6} {\bibfield  {journal}
  {\bibinfo  {journal} {J. Phys. A: Math. Theor.}\ }\textbf {\bibinfo {volume}
  {50}},\ \bibinfo {pages} {453001} (\bibinfo {year} {2017})}\BibitemShut
  {NoStop}%
\bibitem [{\citenamefont {Deffner}(2017)}]{NewJPhys_19_103018_2017}%
  \BibitemOpen
  \bibfield  {author} {\bibinfo {author} {\bibfnamefont {S.}~\bibnamefont
  {Deffner}},\ }\bibfield  {title} {\enquote {\bibinfo {title} {Geometric
  quantum speed limits: a case for {W}igner phase space},}\ }\href {\doibase
  10.1088/1367-2630/aa83dc} {\bibfield  {journal} {\bibinfo  {journal} {New J.
  Phys.}\ }\textbf {\bibinfo {volume} {19}},\ \bibinfo {pages} {103018}
  (\bibinfo {year} {2017})}\BibitemShut {NoStop}%
\bibitem [{\citenamefont {Rosal}\ \emph {et~al.}(2023)\citenamefont {Rosal},
  \citenamefont {Pires},\ and\ \citenamefont
  {Soares-Pinto}}]{arXiv:2312.00533}%
  \BibitemOpen
  \bibfield  {author} {\bibinfo {author} {\bibfnamefont {A.~J.~B.}\
  \bibnamefont {Rosal}}, \bibinfo {author} {\bibfnamefont {D.~P.}\ \bibnamefont
  {Pires}}, \ and\ \bibinfo {author} {\bibfnamefont {D.~O.}\ \bibnamefont
  {Soares-Pinto}},\ }\bibfield  {title} {\enquote {\bibinfo {title} {Quantum
  {S}peed {L}imits based on {S}chatten norms},}\ }\href
  {https://arxiv.org/abs/2312.00533} {\bibfield  {journal} {\bibinfo  {journal}
  {arXiv:2312.00533}\ } (\bibinfo {year} {2023})}\BibitemShut {NoStop}%
\bibitem [{\citenamefont {Wang}\ and\ \citenamefont
  {Qiu}(2024)}]{arXiv:2401.01746}%
  \BibitemOpen
  \bibfield  {author} {\bibinfo {author} {\bibfnamefont {H.}~\bibnamefont
  {Wang}}\ and\ \bibinfo {author} {\bibfnamefont {X.}~\bibnamefont {Qiu}},\
  }\bibfield  {title} {\enquote {\bibinfo {title} {Generalized {C}oherent
  {Q}uantum {S}peed {L}imits},}\ }\href {https://arxiv.org/abs/2401.01746}
  {\bibfield  {journal} {\bibinfo  {journal} {arXiv:2401.01746}\ } (\bibinfo
  {year} {2024})}\BibitemShut {NoStop}%
\bibitem [{\citenamefont {Garc\'{\i}a-Pintos}\ \emph
  {et~al.}(2022)\citenamefont {Garc\'{\i}a-Pintos}, \citenamefont {Nicholson},
  \citenamefont {Green}, \citenamefont {del Campo},\ and\ \citenamefont
  {Gorshkov}}]{PhysRevX.12.011038}%
  \BibitemOpen
  \bibfield  {author} {\bibinfo {author} {\bibfnamefont {L.~P.}\ \bibnamefont
  {Garc\'{\i}a-Pintos}}, \bibinfo {author} {\bibfnamefont {S.~B.}\ \bibnamefont
  {Nicholson}}, \bibinfo {author} {\bibfnamefont {J.~R.}\ \bibnamefont
  {Green}}, \bibinfo {author} {\bibfnamefont {A.}~\bibnamefont {del Campo}}, \
  and\ \bibinfo {author} {\bibfnamefont {A.~V.}\ \bibnamefont {Gorshkov}},\
  }\bibfield  {title} {\enquote {\bibinfo {title} {Unifying {Q}uantum and
  {C}lassical {S}peed {L}imits on {O}bservables},}\ }\href {\doibase
  10.1103/PhysRevX.12.011038} {\bibfield  {journal} {\bibinfo  {journal} {Phys.
  Rev. X}\ }\textbf {\bibinfo {volume} {12}},\ \bibinfo {pages} {011038}
  (\bibinfo {year} {2022})}\BibitemShut {NoStop}%
\bibitem [{\citenamefont {Hamazaki}(2022)}]{PRXQuantum.3.020319}%
  \BibitemOpen
  \bibfield  {author} {\bibinfo {author} {\bibfnamefont {R.}~\bibnamefont
  {Hamazaki}},\ }\bibfield  {title} {\enquote {\bibinfo {title} {Speed {L}imits
  for {M}acroscopic {T}ransitions},}\ }\href {\doibase
  10.1103/PRXQuantum.3.020319} {\bibfield  {journal} {\bibinfo  {journal} {PRX
  Quantum}\ }\textbf {\bibinfo {volume} {3}},\ \bibinfo {pages} {020319}
  (\bibinfo {year} {2022})}\BibitemShut {NoStop}%
\bibitem [{\citenamefont {Lan}\ \emph {et~al.}(2022)\citenamefont {Lan},
  \citenamefont {Xie},\ and\ \citenamefont {Cai}}]{NJPhys_24_055003_2022}%
  \BibitemOpen
  \bibfield  {author} {\bibinfo {author} {\bibfnamefont {K.}~\bibnamefont
  {Lan}}, \bibinfo {author} {\bibfnamefont {S.}~\bibnamefont {Xie}}, \ and\
  \bibinfo {author} {\bibfnamefont {X.}~\bibnamefont {Cai}},\ }\bibfield
  {title} {\enquote {\bibinfo {title} {Geometric quantum speed limits for
  {M}arkovian dynamics in open quantum systems},}\ }\href {\doibase
  10.1088/1367-2630/ac696b} {\bibfield  {journal} {\bibinfo  {journal} {New J.
  Phys.}\ }\textbf {\bibinfo {volume} {24}},\ \bibinfo {pages} {055003}
  (\bibinfo {year} {2022})}\BibitemShut {NoStop}%
\bibitem [{\citenamefont {Xu}\ and\ \citenamefont {Zhu}(2014)}]{Xu2014}%
  \BibitemOpen
  \bibfield  {author} {\bibinfo {author} {\bibfnamefont {Z.-Y.}\ \bibnamefont
  {Xu}}\ and\ \bibinfo {author} {\bibfnamefont {S.-Q.}\ \bibnamefont {Zhu}},\
  }\bibfield  {title} {\enquote {\bibinfo {title} {Quantum {S}peed {L}imit of a
  {P}hoton under {N}on-{M}arkovian {D}ynamics},}\ }\href {\doibase
  10.1088/0256-307x/31/2/020301} {\bibfield  {journal} {\bibinfo  {journal}
  {Chinese Phys. Lett.}\ }\textbf {\bibinfo {volume} {31}},\ \bibinfo {pages}
  {020301} (\bibinfo {year} {2014})}\BibitemShut {NoStop}%
\bibitem [{\citenamefont {Sun}\ \emph {et~al.}(2015)\citenamefont {Sun},
  \citenamefont {Liu}, \citenamefont {Ma},\ and\ \citenamefont
  {Wang}}]{Sun2015}%
  \BibitemOpen
  \bibfield  {author} {\bibinfo {author} {\bibfnamefont {Z.}~\bibnamefont
  {Sun}}, \bibinfo {author} {\bibfnamefont {J.}~\bibnamefont {Liu}}, \bibinfo
  {author} {\bibfnamefont {J.}~\bibnamefont {Ma}}, \ and\ \bibinfo {author}
  {\bibfnamefont {X.}~\bibnamefont {Wang}},\ }\bibfield  {title} {\enquote
  {\bibinfo {title} {Quantum speed limits in open systems: {N}on-{M}arkovian
  dynamics without rotating-wave approximation},}\ }\href {\doibase
  10.1038/srep08444} {\bibfield  {journal} {\bibinfo  {journal} {Sci. Rep.}\
  }\textbf {\bibinfo {volume} {5}},\ \bibinfo {pages} {8444} (\bibinfo {year}
  {2015})}\BibitemShut {NoStop}%
\bibitem [{\citenamefont {Meng}\ \emph {et~al.}(2015)\citenamefont {Meng},
  \citenamefont {Wu},\ and\ \citenamefont {Guo}}]{Meng2015}%
  \BibitemOpen
  \bibfield  {author} {\bibinfo {author} {\bibfnamefont {X.}~\bibnamefont
  {Meng}}, \bibinfo {author} {\bibfnamefont {C.}~\bibnamefont {Wu}}, \ and\
  \bibinfo {author} {\bibfnamefont {H.}~\bibnamefont {Guo}},\ }\bibfield
  {title} {\enquote {\bibinfo {title} {Minimal evolution time and quantum speed
  limit of non-{M}arkovian open systems},}\ }\href {\doibase 10.1038/srep16357}
  {\bibfield  {journal} {\bibinfo  {journal} {Sci. Rep.}\ }\textbf {\bibinfo
  {volume} {5}},\ \bibinfo {pages} {16357} (\bibinfo {year}
  {2015})}\BibitemShut {NoStop}%
\bibitem [{\citenamefont {Mirkin}\ \emph {et~al.}(2016)\citenamefont {Mirkin},
  \citenamefont {Toscano},\ and\ \citenamefont {Wisniacki}}]{Nicolas2016}%
  \BibitemOpen
  \bibfield  {author} {\bibinfo {author} {\bibfnamefont {N.}~\bibnamefont
  {Mirkin}}, \bibinfo {author} {\bibfnamefont {F.}~\bibnamefont {Toscano}}, \
  and\ \bibinfo {author} {\bibfnamefont {D.~A.}\ \bibnamefont {Wisniacki}},\
  }\bibfield  {title} {\enquote {\bibinfo {title} {Quantum-speed-limit bounds
  in an open quantum evolution},}\ }\href {\doibase 10.1103/PhysRevA.94.052125}
  {\bibfield  {journal} {\bibinfo  {journal} {Phys. Rev. A}\ }\textbf {\bibinfo
  {volume} {94}},\ \bibinfo {pages} {052125} (\bibinfo {year}
  {2016})}\BibitemShut {NoStop}%
\bibitem [{\citenamefont {Zhang}\ \emph {et~al.}(2016)\citenamefont {Zhang},
  \citenamefont {Xia},\ and\ \citenamefont {Fan}}]{Zhang2016}%
  \BibitemOpen
  \bibfield  {author} {\bibinfo {author} {\bibfnamefont {Y.-J.}\ \bibnamefont
  {Zhang}}, \bibinfo {author} {\bibfnamefont {Y.-J.}\ \bibnamefont {Xia}}, \
  and\ \bibinfo {author} {\bibfnamefont {H.}~\bibnamefont {Fan}},\ }\bibfield
  {title} {\enquote {\bibinfo {title} {Control of quantum dynamics:
  {N}on-{M}arkovianity and the speedup of the open system evolution},}\ }\href
  {\doibase 10.1209/0295-5075/116/30001} {\bibfield  {journal} {\bibinfo
  {journal} {{EPL} (Europhysics Letters)}\ }\textbf {\bibinfo {volume} {116}},\
  \bibinfo {pages} {30001} (\bibinfo {year} {2016})}\BibitemShut {NoStop}%
\bibitem [{\citenamefont {Rivas}\ \emph {et~al.}(2014)\citenamefont {Rivas},
  \citenamefont {Huelga},\ and\ \citenamefont {Plenio}}]{Rivas_2014}%
  \BibitemOpen
  \bibfield  {author} {\bibinfo {author} {\bibfnamefont {A.}~\bibnamefont
  {Rivas}}, \bibinfo {author} {\bibfnamefont {S.~F.}\ \bibnamefont {Huelga}}, \
  and\ \bibinfo {author} {\bibfnamefont {M.~B.}\ \bibnamefont {Plenio}},\
  }\bibfield  {title} {\enquote {\bibinfo {title} {Quantum non-{M}arkovianity:
  characterization, quantification and detection},}\ }\href {\doibase
  10.1088/0034-4885/77/9/094001} {\bibfield  {journal} {\bibinfo  {journal}
  {Rep. Prog. Phys.}\ }\textbf {\bibinfo {volume} {77}},\ \bibinfo {pages}
  {094001} (\bibinfo {year} {2014})}\BibitemShut {NoStop}%
\bibitem [{\citenamefont {Jing}\ \emph {et~al.}(2016)\citenamefont {Jing},
  \citenamefont {Wu},\ and\ \citenamefont {del Campo}}]{SciRep_6_38149_2016}%
  \BibitemOpen
  \bibfield  {author} {\bibinfo {author} {\bibfnamefont {J.}~\bibnamefont
  {Jing}}, \bibinfo {author} {\bibfnamefont {L.-A.}\ \bibnamefont {Wu}}, \ and\
  \bibinfo {author} {\bibfnamefont {A.}~\bibnamefont {del Campo}},\ }\bibfield
  {title} {\enquote {\bibinfo {title} {Fundamental {S}peed {L}imits to the
  {G}eneration of {Q}uantumness},}\ }\href {\doibase 10.1038/srep38149}
  {\bibfield  {journal} {\bibinfo  {journal} {Sci. Rep.}\ }\textbf {\bibinfo
  {volume} {6}},\ \bibinfo {pages} {38149} (\bibinfo {year}
  {2016})}\BibitemShut {NoStop}%
\bibitem [{\citenamefont {Xu}\ \emph {et~al.}(2014)\citenamefont {Xu},
  \citenamefont {Luo}, \citenamefont {Yang}, \citenamefont {Liu},\ and\
  \citenamefont {Zhu}}]{PhysRevA.89.012307}%
  \BibitemOpen
  \bibfield  {author} {\bibinfo {author} {\bibfnamefont {Z.-Y.}\ \bibnamefont
  {Xu}}, \bibinfo {author} {\bibfnamefont {S.}~\bibnamefont {Luo}}, \bibinfo
  {author} {\bibfnamefont {W.~L.}\ \bibnamefont {Yang}}, \bibinfo {author}
  {\bibfnamefont {C.}~\bibnamefont {Liu}}, \ and\ \bibinfo {author}
  {\bibfnamefont {S.}~\bibnamefont {Zhu}},\ }\bibfield  {title} {\enquote
  {\bibinfo {title} {Quantum speedup in a memory environment},}\ }\href
  {\doibase 10.1103/PhysRevA.89.012307} {\bibfield  {journal} {\bibinfo
  {journal} {Phys. Rev. A}\ }\textbf {\bibinfo {volume} {89}},\ \bibinfo
  {pages} {012307} (\bibinfo {year} {2014})}\BibitemShut {NoStop}%
\bibitem [{\citenamefont {Cianciaruso}\ \emph {et~al.}(2017)\citenamefont
  {Cianciaruso}, \citenamefont {Maniscalco},\ and\ \citenamefont
  {Adesso}}]{Cianciaruso2017}%
  \BibitemOpen
  \bibfield  {author} {\bibinfo {author} {\bibfnamefont {M.}~\bibnamefont
  {Cianciaruso}}, \bibinfo {author} {\bibfnamefont {S.}~\bibnamefont
  {Maniscalco}}, \ and\ \bibinfo {author} {\bibfnamefont {G.}~\bibnamefont
  {Adesso}},\ }\bibfield  {title} {\enquote {\bibinfo {title} {Role of
  non-{M}arkovianity and backflow of information in the speed of quantum
  evolution},}\ }\href {\doibase 10.1103/PhysRevA.96.012105} {\bibfield
  {journal} {\bibinfo  {journal} {Phys. Rev. A}\ }\textbf {\bibinfo {volume}
  {96}},\ \bibinfo {pages} {012105} (\bibinfo {year} {2017})}\BibitemShut
  {NoStop}%
\bibitem [{\citenamefont {Teittinen}\ \emph {et~al.}(2019)\citenamefont
  {Teittinen}, \citenamefont {Lyyra},\ and\ \citenamefont
  {Maniscalco}}]{Teittinen2019}%
  \BibitemOpen
  \bibfield  {author} {\bibinfo {author} {\bibfnamefont {J.}~\bibnamefont
  {Teittinen}}, \bibinfo {author} {\bibfnamefont {H.}~\bibnamefont {Lyyra}}, \
  and\ \bibinfo {author} {\bibfnamefont {S.}~\bibnamefont {Maniscalco}},\
  }\bibfield  {title} {\enquote {\bibinfo {title} {There is no general
  connection between the quantum speed limit and non-{M}arkovianity},}\ }\href
  {\doibase 10.1088/1367-2630/ab59fe} {\bibfield  {journal} {\bibinfo
  {journal} {New J. Phys.}\ }\textbf {\bibinfo {volume} {21}},\ \bibinfo
  {pages} {123041} (\bibinfo {year} {2019})}\BibitemShut {NoStop}%
\bibitem [{\citenamefont {Teittinen}\ and\ \citenamefont
  {Maniscalco}(2021)}]{Entropy_23_331_2021}%
  \BibitemOpen
  \bibfield  {author} {\bibinfo {author} {\bibfnamefont {J.}~\bibnamefont
  {Teittinen}}\ and\ \bibinfo {author} {\bibfnamefont {S.}~\bibnamefont
  {Maniscalco}},\ }\bibfield  {title} {\enquote {\bibinfo {title} {Quantum
  {S}peed {L}imit and {D}ivisibility of the {D}ynamical {M}ap},}\ }\href
  {\doibase 10.3390/e23030331} {\bibfield  {journal} {\bibinfo  {journal}
  {Entropy}\ }\textbf {\bibinfo {volume} {23}},\ \bibinfo {pages} {331}
  (\bibinfo {year} {2021})}\BibitemShut {NoStop}%
\bibitem [{\citenamefont {Ness}\ \emph {et~al.}(2021)\citenamefont {Ness},
  \citenamefont {Lam}, \citenamefont {Alt}, \citenamefont {Meschede},
  \citenamefont {Sagi},\ and\ \citenamefont {Alberti}}]{sciadvabj91192021}%
  \BibitemOpen
  \bibfield  {author} {\bibinfo {author} {\bibfnamefont {G.}~\bibnamefont
  {Ness}}, \bibinfo {author} {\bibfnamefont {M.~R.}\ \bibnamefont {Lam}},
  \bibinfo {author} {\bibfnamefont {W.}~\bibnamefont {Alt}}, \bibinfo {author}
  {\bibfnamefont {D.}~\bibnamefont {Meschede}}, \bibinfo {author}
  {\bibfnamefont {Y.}~\bibnamefont {Sagi}}, \ and\ \bibinfo {author}
  {\bibfnamefont {A.}~\bibnamefont {Alberti}},\ }\bibfield  {title} {\enquote
  {\bibinfo {title} {Observing crossover between quantum speed limits},}\
  }\href {\doibase 10.1126/sciadv.abj9119} {\bibfield  {journal} {\bibinfo
  {journal} {Sci. Adv.}\ }\textbf {\bibinfo {volume} {7}},\ \bibinfo {pages}
  {9119} (\bibinfo {year} {2021})}\BibitemShut {NoStop}%
\bibitem [{\citenamefont {Villamizar}\ \emph {et~al.}(2018)\citenamefont
  {Villamizar}, \citenamefont {Duzzioni}, \citenamefont {Leal},\ and\
  \citenamefont {Auccaise}}]{PhysRevA.97.052125}%
  \BibitemOpen
  \bibfield  {author} {\bibinfo {author} {\bibfnamefont {D.~V.}\ \bibnamefont
  {Villamizar}}, \bibinfo {author} {\bibfnamefont {E.~I.}\ \bibnamefont
  {Duzzioni}}, \bibinfo {author} {\bibfnamefont {A.~C.~S.}\ \bibnamefont
  {Leal}}, \ and\ \bibinfo {author} {\bibfnamefont {R.}~\bibnamefont
  {Auccaise}},\ }\bibfield  {title} {\enquote {\bibinfo {title} {Estimating the
  time evolution of {NMR} systems via a quantum-speed-limit--like
  expression},}\ }\href {\doibase 10.1103/PhysRevA.97.052125} {\bibfield
  {journal} {\bibinfo  {journal} {Phys. Rev. A}\ }\textbf {\bibinfo {volume}
  {97}},\ \bibinfo {pages} {052125} (\bibinfo {year} {2018})}\BibitemShut
  {NoStop}%
\bibitem [{\citenamefont {Kondo}\ \emph {et~al.}(2016)\citenamefont {Kondo},
  \citenamefont {Matsuzaki}, \citenamefont {Matsushima},\ and\ \citenamefont
  {Filgueiras}}]{KondoNJP}%
  \BibitemOpen
  \bibfield  {author} {\bibinfo {author} {\bibfnamefont {Y.}~\bibnamefont
  {Kondo}}, \bibinfo {author} {\bibfnamefont {Y.}~\bibnamefont {Matsuzaki}},
  \bibinfo {author} {\bibfnamefont {K.}~\bibnamefont {Matsushima}}, \ and\
  \bibinfo {author} {\bibfnamefont {J.~G.}\ \bibnamefont {Filgueiras}},\
  }\bibfield  {title} {\enquote {\bibinfo {title} {Using the quantum {Z}eno
  effect for suppression of decoherence},}\ }\href
  {http://stacks.iop.org/1367-2630/18/i=1/a=013033} {\bibfield  {journal}
  {\bibinfo  {journal} {New J. Phys.}\ }\textbf {\bibinfo {volume} {18}},\
  \bibinfo {pages} {013033} (\bibinfo {year} {2016})}\BibitemShut {NoStop}%
\bibitem [{\citenamefont {Ho}\ \emph {et~al.}(2020)\citenamefont {Ho},
  \citenamefont {Matsuzaki}, \citenamefont {Matsuzaki},\ and\ \citenamefont
  {Kondo}}]{Kondo2020}%
  \BibitemOpen
  \bibfield  {author} {\bibinfo {author} {\bibfnamefont {L.~B.}\ \bibnamefont
  {Ho}}, \bibinfo {author} {\bibfnamefont {Y.}~\bibnamefont {Matsuzaki}},
  \bibinfo {author} {\bibfnamefont {M.}~\bibnamefont {Matsuzaki}}, \ and\
  \bibinfo {author} {\bibfnamefont {Y.}~\bibnamefont {Kondo}},\ }\bibfield
  {title} {\enquote {\bibinfo {title} {Nuclear {M}agnetic {R}esonance {M}odel
  of an {E}ntangled {S}ensor under {N}oise},}\ }\href {\doibase
  10.7566/JPSJ.89.054001} {\bibfield  {journal} {\bibinfo  {journal} {J. Phys.
  Soc. Japan}\ }\textbf {\bibinfo {volume} {89}},\ \bibinfo {pages} {054001}
  (\bibinfo {year} {2020})}\BibitemShut {NoStop}%
\bibitem [{\citenamefont {Iwakura}\ \emph {et~al.}(2017)\citenamefont
  {Iwakura}, \citenamefont {Matsuzaki},\ and\ \citenamefont
  {Kondo}}]{KondoPRA}%
  \BibitemOpen
  \bibfield  {author} {\bibinfo {author} {\bibfnamefont {A.}~\bibnamefont
  {Iwakura}}, \bibinfo {author} {\bibfnamefont {Y.}~\bibnamefont {Matsuzaki}},
  \ and\ \bibinfo {author} {\bibfnamefont {Y.}~\bibnamefont {Kondo}},\
  }\bibfield  {title} {\enquote {\bibinfo {title} {Engineered noisy environment
  for studying decoherence},}\ }\href {\doibase 10.1103/PhysRevA.96.032303}
  {\bibfield  {journal} {\bibinfo  {journal} {Phys. Rev. A}\ }\textbf {\bibinfo
  {volume} {96}},\ \bibinfo {pages} {032303} (\bibinfo {year}
  {2017})}\BibitemShut {NoStop}%
\bibitem [{\citenamefont {Abragam}(1961)}]{abragam}%
  \BibitemOpen
  \bibfield  {author} {\bibinfo {author} {\bibfnamefont {A.}~\bibnamefont
  {Abragam}},\ }\href {https://books.google.com.br/books?id=9M8U\_JK7K54C}
  {\emph {\bibinfo {title} {The {P}rinciples of {N}uclear {M}agnetism}}},\
  International series of monographs on physics\ (\bibinfo  {publisher}
  {Clarendon Press},\ \bibinfo {year} {1961})\BibitemShut {NoStop}%
\bibitem [{\citenamefont {Ho}\ \emph {et~al.}(2019)\citenamefont {Ho},
  \citenamefont {Matsuzaki}, \citenamefont {Matsuzaki},\ and\ \citenamefont
  {Kondo}}]{Ho_2019}%
  \BibitemOpen
  \bibfield  {author} {\bibinfo {author} {\bibfnamefont {L.~B.}\ \bibnamefont
  {Ho}}, \bibinfo {author} {\bibfnamefont {Y.}~\bibnamefont {Matsuzaki}},
  \bibinfo {author} {\bibfnamefont {M.}~\bibnamefont {Matsuzaki}}, \ and\
  \bibinfo {author} {\bibfnamefont {Y.}~\bibnamefont {Kondo}},\ }\bibfield
  {title} {\enquote {\bibinfo {title} {Realization of controllable open system
  with {NMR}},}\ }\href {\doibase 10.1088/1367-2630/ab3a25} {\bibfield
  {journal} {\bibinfo  {journal} {New J. Phys.}\ }\textbf {\bibinfo {volume}
  {21}},\ \bibinfo {pages} {093008} (\bibinfo {year} {2019})}\BibitemShut
  {NoStop}%
\bibitem [{Sup()}]{SupMat}%
  \BibitemOpen
  \href@noop {} {}\bibinfo {note} {Supplementary Materials at [URL will be
  inserted by publisher] for technical details about the experiment and sample
  preparation, and details on the derivation of geometric QSLs}\BibitemShut
  {NoStop}%
\bibitem [{\citenamefont {Zhou}\ \emph {et~al.}(2007)\citenamefont {Zhou},
  \citenamefont {Kummerle}, \citenamefont {Qiu}, \citenamefont {Redwine},
  \citenamefont {Cong}, \citenamefont {Taha}, \citenamefont {Baugh},\ and\
  \citenamefont {Winniford}}]{Waltz64}%
  \BibitemOpen
  \bibfield  {author} {\bibinfo {author} {\bibfnamefont {Z.}~\bibnamefont
  {Zhou}}, \bibinfo {author} {\bibfnamefont {R.}~\bibnamefont {Kummerle}},
  \bibinfo {author} {\bibfnamefont {X.}~\bibnamefont {Qiu}}, \bibinfo {author}
  {\bibfnamefont {D.}~\bibnamefont {Redwine}}, \bibinfo {author} {\bibfnamefont
  {R.}~\bibnamefont {Cong}}, \bibinfo {author} {\bibfnamefont {A.}~\bibnamefont
  {Taha}}, \bibinfo {author} {\bibfnamefont {D.}~\bibnamefont {Baugh}}, \ and\
  \bibinfo {author} {\bibfnamefont {B.}~\bibnamefont {Winniford}},\ }\bibfield
  {title} {\enquote {\bibinfo {title} {A new decoupling method for accurate
  quantification of polyethylene copolymer composition and triad sequence
  distribution with $^{13}${C} {NMR}},}\ }\href {\doibase
  10.1016/j.jmr.2007.05.005} {\bibfield  {journal} {\bibinfo  {journal} {J.
  Magn. Reson.}\ }\textbf {\bibinfo {volume} {187}},\ \bibinfo {pages} {225}
  (\bibinfo {year} {2007})}\BibitemShut {NoStop}%
\bibitem [{\citenamefont {Morozova}\ and\ \citenamefont
  {\v{C}encov}(1991)}]{MC-paper}%
  \BibitemOpen
  \bibfield  {author} {\bibinfo {author} {\bibfnamefont {E.~A.}\ \bibnamefont
  {Morozova}}\ and\ \bibinfo {author} {\bibfnamefont {N.~N.}\ \bibnamefont
  {\v{C}encov}},\ }\bibfield  {title} {\enquote {\bibinfo {title} {Markov
  {I}nvariant {G}eometry on {M}anifolds of {S}tates},}\ }\href {\doibase
  10.1007/BF01095975} {\bibfield  {journal} {\bibinfo  {journal} {J. Sov.
  Math.}\ }\textbf {\bibinfo {volume} {56}},\ \bibinfo {pages} {2648} (\bibinfo
  {year} {1991})}\BibitemShut {NoStop}%
\bibitem [{\citenamefont {Petz}(1996)}]{Petz-paper}%
  \BibitemOpen
  \bibfield  {author} {\bibinfo {author} {\bibfnamefont {D.}~\bibnamefont
  {Petz}},\ }\bibfield  {title} {\enquote {\bibinfo {title} {Monotone metrics
  on matrix spaces},}\ }\href {\doibase 10.1016/0024-3795(94)00211-8}
  {\bibfield  {journal} {\bibinfo  {journal} {Linear Algebra Appl.}\ }\textbf
  {\bibinfo {volume} {244}},\ \bibinfo {pages} {81} (\bibinfo {year}
  {1996})}\BibitemShut {NoStop}%
\bibitem [{\citenamefont {Bengtsson}\ and\ \citenamefont
  {\.{Z}yczkowski}(2006)}]{Ingemar_Bengtsson_Zyczkowski}%
  \BibitemOpen
  \bibfield  {author} {\bibinfo {author} {\bibfnamefont {I.}~\bibnamefont
  {Bengtsson}}\ and\ \bibinfo {author} {\bibfnamefont {K.}~\bibnamefont
  {\.{Z}yczkowski}},\ }\href {\doibase 10.1017/CBO9780511535048} {\emph
  {\bibinfo {title} {Geometry of {Q}uantum {S}tates: {A}n {I}ntroduction to
  {Q}uantum {E}ntanglement}}}\ (\bibinfo  {publisher} {Cambridge University
  Press},\ \bibinfo {address} {Cambridge},\ \bibinfo {year} {2006})\BibitemShut
  {NoStop}%
\bibitem [{\citenamefont {Nielsen}\ and\ \citenamefont
  {Chuang}(2010)}]{Nielsen_Chuang_infor_geom}%
  \BibitemOpen
  \bibfield  {author} {\bibinfo {author} {\bibfnamefont {M.}~\bibnamefont
  {Nielsen}}\ and\ \bibinfo {author} {\bibfnamefont {I.~L.}\ \bibnamefont
  {Chuang}},\ }\href {\doibase 10.1017/CBO9780511976667} {\emph {\bibinfo
  {title} {Quantum {C}omputation and {Q}uantum {I}nformation: 10th
  {A}nniversary {E}dition}}}\ (\bibinfo  {publisher} {Cambridge University
  Press},\ \bibinfo {address} {Cambridge},\ \bibinfo {year} {2010})\BibitemShut
  {NoStop}%
\bibitem [{\citenamefont {Anandan}\ and\ \citenamefont
  {Aharonov}(1990)}]{PhysRevLett.65.1697}%
  \BibitemOpen
  \bibfield  {author} {\bibinfo {author} {\bibfnamefont {J.}~\bibnamefont
  {Anandan}}\ and\ \bibinfo {author} {\bibfnamefont {Y.}~\bibnamefont
  {Aharonov}},\ }\bibfield  {title} {\enquote {\bibinfo {title} {Geometry of
  quantum evolution},}\ }\href {\doibase 10.1103/PhysRevLett.65.1697}
  {\bibfield  {journal} {\bibinfo  {journal} {Phys. Rev. Lett.}\ }\textbf
  {\bibinfo {volume} {65}},\ \bibinfo {pages} {1697} (\bibinfo {year}
  {1990})}\BibitemShut {NoStop}%
\bibitem [{\citenamefont {Breuer}\ \emph {et~al.}(2009)\citenamefont {Breuer},
  \citenamefont {Laine},\ and\ \citenamefont {Piilo}}]{Breuer2009}%
  \BibitemOpen
  \bibfield  {author} {\bibinfo {author} {\bibfnamefont {H.~P.}\ \bibnamefont
  {Breuer}}, \bibinfo {author} {\bibfnamefont {E.~M.}\ \bibnamefont {Laine}}, \
  and\ \bibinfo {author} {\bibfnamefont {J.}~\bibnamefont {Piilo}},\ }\bibfield
   {title} {\enquote {\bibinfo {title} {Measure for the {D}egree of
  {N}on-{M}arkovian {B}ehavior of {Q}uantum {P}rocesses in {O}pen {S}ystems},}\
  }\href {\doibase 10.1103/PhysRevLett.103.210401} {\bibfield  {journal}
  {\bibinfo  {journal} {Phys. Rev. Lett.}\ }\textbf {\bibinfo {volume} {103}},\
  \bibinfo {pages} {210401} (\bibinfo {year} {2009})}\BibitemShut {NoStop}%
\bibitem [{\citenamefont {Rivas}\ \emph {et~al.}(2010)\citenamefont {Rivas},
  \citenamefont {Huelga},\ and\ \citenamefont
  {Plenio}}]{PhysRevLett.105.050403}%
  \BibitemOpen
  \bibfield  {author} {\bibinfo {author} {\bibfnamefont {A.}~\bibnamefont
  {Rivas}}, \bibinfo {author} {\bibfnamefont {S.~F.}\ \bibnamefont {Huelga}}, \
  and\ \bibinfo {author} {\bibfnamefont {M.~B.}\ \bibnamefont {Plenio}},\
  }\bibfield  {title} {\enquote {\bibinfo {title} {Entanglement and
  {N}on-{M}arkovianity of {Q}uantum {E}volutions},}\ }\href {\doibase
  10.1103/PhysRevLett.105.050403} {\bibfield  {journal} {\bibinfo  {journal}
  {Phys. Rev. Lett.}\ }\textbf {\bibinfo {volume} {105}},\ \bibinfo {pages}
  {050403} (\bibinfo {year} {2010})}\BibitemShut {NoStop}%
\bibitem [{\citenamefont {Luo}\ \emph {et~al.}(2012)\citenamefont {Luo},
  \citenamefont {Fu},\ and\ \citenamefont {Song}}]{PhysRevA.86.044101}%
  \BibitemOpen
  \bibfield  {author} {\bibinfo {author} {\bibfnamefont {S.}~\bibnamefont
  {Luo}}, \bibinfo {author} {\bibfnamefont {S.}~\bibnamefont {Fu}}, \ and\
  \bibinfo {author} {\bibfnamefont {H.}~\bibnamefont {Song}},\ }\bibfield
  {title} {\enquote {\bibinfo {title} {Quantifying non-{M}arkovianity via
  correlations},}\ }\href {\doibase 10.1103/PhysRevA.86.044101} {\bibfield
  {journal} {\bibinfo  {journal} {Phys. Rev. A}\ }\textbf {\bibinfo {volume}
  {86}},\ \bibinfo {pages} {044101} (\bibinfo {year} {2012})}\BibitemShut
  {NoStop}%
\bibitem [{\citenamefont {Rivas}(2017)}]{PhysRevA.95.042104}%
  \BibitemOpen
  \bibfield  {author} {\bibinfo {author} {\bibfnamefont {A.}~\bibnamefont
  {Rivas}},\ }\bibfield  {title} {\enquote {\bibinfo {title} {Refined
  weak-coupling limit: {C}oherence, entanglement, and non-{M}arkovianity},}\
  }\href {\doibase 10.1103/PhysRevA.95.042104} {\bibfield  {journal} {\bibinfo
  {journal} {Phys. Rev. A}\ }\textbf {\bibinfo {volume} {95}},\ \bibinfo
  {pages} {042104} (\bibinfo {year} {2017})}\BibitemShut {NoStop}%
\bibitem [{\citenamefont {He}\ \emph {et~al.}(2017)\citenamefont {He},
  \citenamefont {Zeng}, \citenamefont {Li}, \citenamefont {Wang},\ and\
  \citenamefont {Yao}}]{PhysRevA.96.022106}%
  \BibitemOpen
  \bibfield  {author} {\bibinfo {author} {\bibfnamefont {Z.}~\bibnamefont
  {He}}, \bibinfo {author} {\bibfnamefont {H.-S.}\ \bibnamefont {Zeng}},
  \bibinfo {author} {\bibfnamefont {Y.}~\bibnamefont {Li}}, \bibinfo {author}
  {\bibfnamefont {Q.}~\bibnamefont {Wang}}, \ and\ \bibinfo {author}
  {\bibfnamefont {C.}~\bibnamefont {Yao}},\ }\bibfield  {title} {\enquote
  {\bibinfo {title} {Non-{M}arkovianity measure based on the relative entropy
  of coherence in an extended space},}\ }\href {\doibase
  10.1103/PhysRevA.96.022106} {\bibfield  {journal} {\bibinfo  {journal} {Phys.
  Rev. A}\ }\textbf {\bibinfo {volume} {96}},\ \bibinfo {pages} {022106}
  (\bibinfo {year} {2017})}\BibitemShut {NoStop}%
\bibitem [{\citenamefont {Radhakrishnan}\ \emph {et~al.}(2019)\citenamefont
  {Radhakrishnan}, \citenamefont {Chen}, \citenamefont {Jambulingam},
  \citenamefont {Byrnes},\ and\ \citenamefont {Ali}}]{SciRep_9_2363_2019}%
  \BibitemOpen
  \bibfield  {author} {\bibinfo {author} {\bibfnamefont {C.}~\bibnamefont
  {Radhakrishnan}}, \bibinfo {author} {\bibfnamefont {P.-W.}\ \bibnamefont
  {Chen}}, \bibinfo {author} {\bibfnamefont {S.}~\bibnamefont {Jambulingam}},
  \bibinfo {author} {\bibfnamefont {T.}~\bibnamefont {Byrnes}}, \ and\ \bibinfo
  {author} {\bibfnamefont {Md.~M.}\ \bibnamefont {Ali}},\ }\bibfield  {title}
  {\enquote {\bibinfo {title} {Time dynamics of quantum coherence and monogamy
  in a non-{M}arkovian environment},}\ }\href {\doibase
  10.1038/s41598-019-39027-2} {\bibfield  {journal} {\bibinfo  {journal} {Sci.
  Rep}\ }\textbf {\bibinfo {volume} {9}},\ \bibinfo {pages} {2363} (\bibinfo
  {year} {2019})}\BibitemShut {NoStop}%
\bibitem [{\citenamefont {Chanda}\ and\ \citenamefont
  {Bhattacharya}(2016)}]{AnnPhys_366_1_2016}%
  \BibitemOpen
  \bibfield  {author} {\bibinfo {author} {\bibfnamefont {T.}~\bibnamefont
  {Chanda}}\ and\ \bibinfo {author} {\bibfnamefont {S.}~\bibnamefont
  {Bhattacharya}},\ }\bibfield  {title} {\enquote {\bibinfo {title}
  {Delineating incoherent non-{M}arkovian dynamics using quantum coherence},}\
  }\href {\doibase 10.1016/j.aop.2016.01.004} {\bibfield  {journal} {\bibinfo
  {journal} {Ann. Phys.}\ }\textbf {\bibinfo {volume} {366}},\ \bibinfo {pages}
  {1} (\bibinfo {year} {2016})}\BibitemShut {NoStop}%
\bibitem [{\citenamefont {Du}\ \emph {et~al.}(2020)\citenamefont {Du},
  \citenamefont {Hou}, \citenamefont {Xiang}, \citenamefont {Li}, \citenamefont
  {Guo}, \citenamefont {Dong},\ and\ \citenamefont
  {Nori}}]{npj_Quantum_Inf_6_55_2020}%
  \BibitemOpen
  \bibfield  {author} {\bibinfo {author} {\bibfnamefont {K.-D.}\ \bibnamefont
  {Du}}, \bibinfo {author} {\bibfnamefont {Z.}~\bibnamefont {Hou}}, \bibinfo
  {author} {\bibfnamefont {G.-Y.}\ \bibnamefont {Xiang}}, \bibinfo {author}
  {\bibfnamefont {C.-F.}\ \bibnamefont {Li}}, \bibinfo {author} {\bibfnamefont
  {G.-C.}\ \bibnamefont {Guo}}, \bibinfo {author} {\bibfnamefont
  {D.}~\bibnamefont {Dong}}, \ and\ \bibinfo {author} {\bibfnamefont
  {F.}~\bibnamefont {Nori}},\ }\bibfield  {title} {\enquote {\bibinfo {title}
  {Detecting non-{M}arkovianity via quantified coherence: theory and
  experiments},}\ }\href {\doibase 10.1038/s41534-020-0283-3} {\bibfield
  {journal} {\bibinfo  {journal} {npj {Q}uantum {I}nf.}\ }\textbf {\bibinfo
  {volume} {6}},\ \bibinfo {pages} {55} (\bibinfo {year} {2020})}\BibitemShut
  {NoStop}%
\bibitem [{\citenamefont {Baumgratz}\ \emph {et~al.}(2014)\citenamefont
  {Baumgratz}, \citenamefont {Cramer},\ and\ \citenamefont
  {Plenio}}]{PhysRevLett.113.140401}%
  \BibitemOpen
  \bibfield  {author} {\bibinfo {author} {\bibfnamefont {T.}~\bibnamefont
  {Baumgratz}}, \bibinfo {author} {\bibfnamefont {M.}~\bibnamefont {Cramer}}, \
  and\ \bibinfo {author} {\bibfnamefont {M.~B.}\ \bibnamefont {Plenio}},\
  }\bibfield  {title} {\enquote {\bibinfo {title} {Quantifying {C}oherence},}\
  }\href {\doibase 10.1103/PhysRevLett.113.140401} {\bibfield  {journal}
  {\bibinfo  {journal} {Phys. Rev. Lett.}\ }\textbf {\bibinfo {volume} {113}},\
  \bibinfo {pages} {140401} (\bibinfo {year} {2014})}\BibitemShut {NoStop}%
\bibitem [{\citenamefont {Streltsov}\ \emph {et~al.}(2017)\citenamefont
  {Streltsov}, \citenamefont {Adesso},\ and\ \citenamefont
  {Plenio}}]{RevModPhys.89.041003}%
  \BibitemOpen
  \bibfield  {author} {\bibinfo {author} {\bibfnamefont {A.}~\bibnamefont
  {Streltsov}}, \bibinfo {author} {\bibfnamefont {G.}~\bibnamefont {Adesso}}, \
  and\ \bibinfo {author} {\bibfnamefont {M.~B.}\ \bibnamefont {Plenio}},\
  }\bibfield  {title} {\enquote {\bibinfo {title} {Colloquium: {Q}uantum
  coherence as a resource},}\ }\href {\doibase 10.1103/RevModPhys.89.041003}
  {\bibfield  {journal} {\bibinfo  {journal} {Rev. Mod. Phys.}\ }\textbf
  {\bibinfo {volume} {89}},\ \bibinfo {pages} {041003} (\bibinfo {year}
  {2017})}\BibitemShut {NoStop}%
\bibitem [{\citenamefont {Yu}\ \emph {et~al.}(2022)\citenamefont {Yu},
  \citenamefont {Liu}, \citenamefont {Yang}, \citenamefont {Gong},
  \citenamefont {Cao}, \citenamefont {Zhang}, \citenamefont {Liu},
  \citenamefont {Heyl}, \citenamefont {Ozawa}, \citenamefont {Goldman},\ and\
  \citenamefont {Cai}}]{npjQuantInf_8_56_2022}%
  \BibitemOpen
  \bibfield  {author} {\bibinfo {author} {\bibfnamefont {M.}~\bibnamefont
  {Yu}}, \bibinfo {author} {\bibfnamefont {Y.}~\bibnamefont {Liu}}, \bibinfo
  {author} {\bibfnamefont {P.}~\bibnamefont {Yang}}, \bibinfo {author}
  {\bibfnamefont {M.}~\bibnamefont {Gong}}, \bibinfo {author} {\bibfnamefont
  {Q.}~\bibnamefont {Cao}}, \bibinfo {author} {\bibfnamefont {S.}~\bibnamefont
  {Zhang}}, \bibinfo {author} {\bibfnamefont {H.}~\bibnamefont {Liu}}, \bibinfo
  {author} {\bibfnamefont {M.}~\bibnamefont {Heyl}}, \bibinfo {author}
  {\bibfnamefont {T.}~\bibnamefont {Ozawa}}, \bibinfo {author} {\bibfnamefont
  {N.}~\bibnamefont {Goldman}}, \ and\ \bibinfo {author} {\bibfnamefont
  {J.}~\bibnamefont {Cai}},\ }\bibfield  {title} {\enquote {\bibinfo {title}
  {Quantum {F}isher information measurement and verification of the quantum
  {C}ram\'{e}r-{R}ao bound in a solid-state qubit},}\ }\href {\doibase
  10.1038/s41534-022-00547-x} {\bibfield  {journal} {\bibinfo  {journal} {npj
  Quantum Inf.}\ }\textbf {\bibinfo {volume} {8}},\ \bibinfo {pages} {56}
  (\bibinfo {year} {2022})}\BibitemShut {NoStop}%
\bibitem [{\citenamefont {Zhang}\ \emph {et~al.}(2023)\citenamefont {Zhang},
  \citenamefont {Lu}, \citenamefont {Liu}, \citenamefont {Ding},\ and\
  \citenamefont {Wang}}]{PhysRevA.107.012414}%
  \BibitemOpen
  \bibfield  {author} {\bibinfo {author} {\bibfnamefont {X.}~\bibnamefont
  {Zhang}}, \bibinfo {author} {\bibfnamefont {X.-M.}\ \bibnamefont {Lu}},
  \bibinfo {author} {\bibfnamefont {J.}~\bibnamefont {Liu}}, \bibinfo {author}
  {\bibfnamefont {W.}~\bibnamefont {Ding}}, \ and\ \bibinfo {author}
  {\bibfnamefont {X.}~\bibnamefont {Wang}},\ }\bibfield  {title} {\enquote
  {\bibinfo {title} {Direct measurement of quantum {F}isher information},}\
  }\href {\doibase 10.1103/PhysRevA.107.012414} {\bibfield  {journal} {\bibinfo
   {journal} {Phys. Rev. A}\ }\textbf {\bibinfo {volume} {107}},\ \bibinfo
  {pages} {012414} (\bibinfo {year} {2023})}\BibitemShut {NoStop}%
\bibitem [{\citenamefont {Luo}(2006)}]{PhysRevA.73.022324}%
  \BibitemOpen
  \bibfield  {author} {\bibinfo {author} {\bibfnamefont {S.}~\bibnamefont
  {Luo}},\ }\bibfield  {title} {\enquote {\bibinfo {title} {Quantum uncertainty
  of mixed states based on skew information},}\ }\href {\doibase
  10.1103/PhysRevA.73.022324} {\bibfield  {journal} {\bibinfo  {journal} {Phys.
  Rev. A}\ }\textbf {\bibinfo {volume} {73}},\ \bibinfo {pages} {022324}
  (\bibinfo {year} {2006})}\BibitemShut {NoStop}%
\bibitem [{\citenamefont {T\'oth}\ and\ \citenamefont
  {Petz}(2013)}]{PhysRevA.87.032324}%
  \BibitemOpen
  \bibfield  {author} {\bibinfo {author} {\bibfnamefont {G.}~\bibnamefont
  {T\'oth}}\ and\ \bibinfo {author} {\bibfnamefont {D.}~\bibnamefont {Petz}},\
  }\bibfield  {title} {\enquote {\bibinfo {title} {Extremal properties of the
  variance and the quantum {F}isher information},}\ }\href {\doibase
  10.1103/PhysRevA.87.032324} {\bibfield  {journal} {\bibinfo  {journal} {Phys.
  Rev. A}\ }\textbf {\bibinfo {volume} {87}},\ \bibinfo {pages} {032324}
  (\bibinfo {year} {2013})}\BibitemShut {NoStop}%
\end{thebibliography}


%

\end{document}